%Paper: 9112037
%From: vipul@guinness.ias.edu (Vipul Periwal)
%Date: Mon, 16 Dec 91 18:51:12 EST

% standard tex, all macros included
%
\def\cit#1{[\cite{#1}]}
\def\del{\delta}
\def\th{\theta}
\def\section#1{\bigskip\centerline{#1}\bigskip}
\def\subsection#1{\bigskip\leftline{\ \ \ \ \ #1}\medskip}
\def\bb{$\bullet$}
\def\cit#1{[\cite{#1}]}
\def\tr{{\rm tr}\,}
\def\cz{{\cal Z}}
\def\be{\beta}
\def\al{\alpha}
\def\dif{{\rm d}}

\def\la{\lambda}

\def\ln{{\rm ln}}
\def\ssc{\scriptscriptstyle}

\def\de{\delta}

\def\raro{\rightarrow}

\def\ie{{\it i.e.,}\ }

\def\half{{1\over2}}
\def\quart{{1\over4}}

\def\al{\alpha}

\def\part{\partial}
\font\tif=cmr7 scaled\magstep4
\catcode`@=11
\expandafter\ifx\csname inp@t\endcsname\relax\let\inp@t=\input
\def\input#1 {\expandafter\ifx\csname #1IsLoaded\endcsname\relax
\inp@t#1%
\expandafter\def\csname #1IsLoaded\endcsname{(#1 was previously loaded)}
\else\message{\csname #1IsLoaded\endcsname}\fi}\fi
\catcode`@=12

%%
%%  Font definitions for Computer Modern (CM) Fonts
%%
\font\twelverm=cmr12			\font\twelvei=cmmi12
\font\twelvesy=cmsy10 scaled 1200	\font\twelveex=cmex10 scaled 1200
\font\twelvebf=cmbx12			\font\twelvesl=cmsl12
\font\twelvett=cmtt12			\font\twelveit=cmti12
\font\twelvesc=cmcsc10 scaled 1200	\font\twelvesf=cmss12
\skewchar\twelvei='177			\skewchar\twelvesy='60

%  Define \...point macros to change fonts and spacings consistently

\def\twelvepoint{\normalbaselineskip=12.4pt plus 0.1pt minus 0.1pt
  \abovedisplayskip 12.4pt plus 3pt minus 9pt
  \belowdisplayskip 12.4pt plus 3pt minus 9pt
  \abovedisplayshortskip 0pt plus 3pt
  \belowdisplayshortskip 7.2pt plus 3pt minus 4pt
  \smallskipamount=3.6pt plus1.2pt minus1.2pt
  \medskipamount=7.2pt plus2.4pt minus2.4pt
  \bigskipamount=14.4pt plus4.8pt minus4.8pt
  \def\rm{\fam0\twelverm}          \def\it{\fam\itfam\twelveit}%
  \def\sl{\fam\slfam\twelvesl}     \def\bf{\fam\bffam\twelvebf}%
  \def\mit{\fam 1}                 \def\cal{\fam 2}%
  \def\sc{\twelvesc}		   \def\tt{\twelvett}
  \def\sf{\twelvesf}
  \textfont0=\twelverm   \scriptfont0=\tenrm   \scriptscriptfont0=\sevenrm
  \textfont1=\twelvei    \scriptfont1=\teni    \scriptscriptfont1=\seveni
  \textfont2=\twelvesy   \scriptfont2=\tensy   \scriptscriptfont2=\sevensy
  \textfont3=\twelveex   \scriptfont3=\twelveex  \scriptscriptfont3=\twelveex
  \textfont\itfam=\twelveit
  \textfont\slfam=\twelvesl
  \textfont\bffam=\twelvebf \scriptfont\bffam=\tenbf
  \scriptscriptfont\bffam=\sevenbf
  \normalbaselines\rm}

%	tenpoint

%%
%%	Various internal macros
%%

\def\beginlinemode{\endmode
  \begingroup\parskip=0pt \obeylines\def\\{\par}\def\endmode{\par\endgroup}}
\def\beginparmode{\endmode
  \begingroup \def\endmode{\par\endgroup}}
\let\endmode=\par
{\obeylines\gdef\
{}}
\def\singlespace{\baselineskip=\normalbaselineskip}

\def\oneandahalfspace{\baselineskip=\normalbaselineskip
  \multiply\baselineskip by 3 \divide\baselineskip by 2}
\def\doublespace{\baselineskip=\normalbaselineskip \multiply\baselineskip by 2}

\newcount\firstpageno
\firstpageno=2
%% FOLLOWING LINE CANNOT BE BROKEN BEFORE 80 CHAR
\footline={\ifnum\pageno<\firstpageno{\hfil}\else{\hfil\twelverm\folio\hfil}\fi}
\def\toppageno{\global\footline={\hfil}\global\headline
  ={\ifnum\pageno<\firstpageno{\hfil}\else{\hfil\twelverm\folio\hfil}\fi}}
\let\rawfootnote=\footnote		% We must set the footnote style
\def\footnote#1#2{{\rm\singlespace\parindent=0pt\parskip=0pt
  \rawfootnote{#1}{#2\hfill\vrule height 0pt depth 6pt width 0pt}}}
\def\raggedcenter{\leftskip=4em plus 12em \rightskip=\leftskip
  \parindent=0pt \parfillskip=0pt \spaceskip=.3333em \xspaceskip=.5em
  \pretolerance=9999 \tolerance=9999
  \hyphenpenalty=9999 \exhyphenpenalty=9999 }
\def\dateline{\rightline{\ifcase\month\or
  January\or February\or March\or April\or May\or June\or
  July\or August\or September\or October\or November\or December\fi
  \space\number\year}}
\def\received{\vskip 3pt plus 0.2fill
 \centerline{\sl (Received\space\ifcase\month\or
  January\or February\or March\or April\or May\or June\or
  July\or August\or September\or October\or November\or December\fi
  \qquad, \number\year)}}

%%
%%	Page layout, margins, font and spacing (feel free to change)
%%

\hsize=6.5truein
\hoffset=0pt
%%\hoffset=1truein
\vsize=8.9truein
\voffset=0pt
%%\voffset=1truein
\parskip=\medskipamount
\def\\{\cr}
\twelvepoint		% selects twelvepoint fonts (cf. \tenpoint)
\doublespace		% selects double spacing for main part of paper (cf.
			%	\singlespace, \oneandahalfspace)
\overfullrule=0pt	% delete the nasty little black boxes for overfull box

%%
%%	The user definitions for major parts of a paper (feel free to change)
%%

  %  "Draft", Timestamp

	% Preprint number at upper right of title page

\def\title			%  Title on title page
  {\null\vskip 3pt plus 0.2fill
   \beginlinemode \doublespace \raggedcenter \bf}

\def\author			%  Author(s) name(s)  on title page
  {\vskip 3pt plus 0.2fill \beginlinemode
   \singlespace \raggedcenter\sc}

\def\affil			% Affiliations (can intermix with \author)
  {\vskip 3pt plus 0.1fill \beginlinemode
   \oneandahalfspace \raggedcenter \sl}

\def\abstract			% Begin abstract
  {\vskip 3pt plus 0.3fill \beginparmode
   \baselineskip=16truept ABSTRACT: }

\def\endtitlepage		% End title page, begin body of paper
  {\endpage			% 	This subsumes \body
   \body}
\let\endtopmatter=\endtitlepage

\def\body			% Begin text body;  can be used to end
  {\beginparmode}		% \title, \author, \affil, \abstract,
				% \reference, or \figurecaption modes

\def\head#1{			% Head;  NOTE enclose the text in {}
  \goodbreak\vskip 0.5truein	%  e.g., \head{I. Introduction}
  {\immediate\write16{#1}
   \raggedcenter \uppercase{#1}\par}
   \nobreak\vskip 0.25truein\nobreak}

\def\beginitems{
\par\medskip\bgroup\def\i##1 {\item{##1}}\def\ii##1 {\itemitem{##1}}
\leftskip=36pt\parskip=0pt}
\def\enditems{\par\egroup}

\def\beneathrel#1\under#2{\mathrel{\mathop{#2}\limits_{#1}}}

\def\refto#1{$^{#1}$}		% For references in text as superscript

\def\references			% Begin references -- basic format is Phys Rev
  {\head{References}		% I.e., volume, page, year (space after commas).
   \beginparmode
   \frenchspacing \parindent=0pt \leftskip=1truecm
   \parskip=3pt plus 3pt \baselineskip=17truept %
   \everypar{\hangindent=\parindent}}

\gdef\refis#1{\item{#1.\ }}			% Ref list numbers.

\gdef\journal#1, #2, #3, 1#4#5#6{		% Journal reference.  Comma sets
    {\sl #1~}{\bf #2}, #3 (1#4#5#6)}		% off: name, vol, page, year

\def\endreferences{\body}

\def\figurecaptions		% Begin figure captions
  {\endpage
   \beginparmode
   \head{Figure Captions}
}

\def\endpage			%  Eject a page
  {\vfill\eject}

\def\endpaper			%  Ways to say goodbye
  {\endmode\vfill\supereject}

%%
%%	AmSTeX compatability definitions
%%
%%	To run a TeX file originally intended for AmSTeX, only small changes
%%	should be necessary (I hope).  Use the line \input jnl at the start.
%%	Remove the lines \input amstex, \documentstyle{itpjnl} at the
%%	beginning;  also remove all the page layout stuff (\parindent=1cm,
%%	\hsize=5.28125in etc.)  The page layout is now done automatically.
%%	Also OMIT the qualifier \magnification=1200 when you IMPRINT the
%%	.dvi file.  (\TagsOnRight is harmless, you can take it out or leave
%%	it in.)  I believe most AmSTeX will work with no change.  One problem
%%	is \footnote, which is a little different in that it now needs to
%%	have an explicit asterisk *  (or whatever) included, like this:
%%		\footnote*{Text winds up at bottom of page.}
%%	This is discussed on p. 116 of the TeXbook.  IGNORE the AmSTeX
%%	documentation (if you can call it that);  refer to the TeXbook.
%%
%%	Note that many commands in AmSTeX have their equivalents in the
%%	TeXbook, perhaps with different names and slightly differing
%%	usage. E.g., the old \align in AmSTeX is replaced by \eqalign
%%	(p. 190) and \aligntag is replaced by \eqalignno (p. 192).
%%	\align and \aligntag still work, but I recommend that you use
%%	\eqalign and \eqalignno in documents run under jnl.
%%
%%	See me if you have any problems  -- Doug.
%%

\def\heading				% Heading
  {\vskip 0.5truein plus 0.1truein	% e.g., \heading I. NOTES \endheading
   \beginparmode \def\\{\par} \parskip=0pt \singlespace \raggedcenter}

\def\subheading				% Subheading
  {\vskip 0.25truein plus 0.1truein	% e.g., \subheading{A. The Problem}
   \beginlinemode \singlespace \parskip=0pt \def\\{\par}\raggedcenter}

\def\tag#1$${\eqno(#1)$$}

\def\align#1$${\eqalign{#1}$$}

\def\aligntag#1$${\gdef\tag##1\\{&(##1)\cr}\eqalignno{#1\\}$$
  \gdef\tag##1$${\eqno(##1)$$}}

\def\endaligntag{}

\def\overset #1\to#2{{\mathop{#2}\limits^{#1}}}
\def\underset#1\to#2{{\let\next=#1\mathpalette\undersetpalette#2}}
\def\undersetpalette#1#2{\vtop{\baselineskip0pt
\ialign{$\mathsurround=0pt #1\hfil##\hfil$\crcr#2\crcr\next\crcr}}}

%%
%%	Various little user definitions
%%

\def\ref#1{Ref.~#1}			% 	for inline references
\def\Ref#1{Ref.~#1}			% 	ditto
\def\[#1]{[\cite{#1}]}
\def\cite#1{{#1}}
			% For figure numbers
		% For citation of equation numbers
	%	ditto
			%	ditto
			%	ditto
		%	ditto
\def\(#1){(\call{#1})}
\def\call#1{{#1}}
\def\taghead#1{}
\def\frac#1#2{{#1 \over #2}}
\def\half{{\frac 12}}

\def\12{{1\over2}}

\def\ie{{\it i.e.,\ }}

\def\sla{\raise.15ex\hbox{$/$}\kern-.57em}
\def\leaderfill{\leaders\hbox to 1em{\hss.\hss}\hfill}
\def\twiddle{\lower.9ex\rlap{$\kern-.1em\scriptstyle\sim$}}
\def\bigtwiddle{\lower1.ex\rlap{$\sim$}}
\def\gtwid{\mathrel{\raise.3ex\hbox{$>$\kern-.75em\lower1ex\hbox{$\sim$}}}}
\def\ltwid{\mathrel{\raise.3ex\hbox{$<$\kern-.75em\lower1ex\hbox{$\sim$}}}}
\def\square{\kern1pt\vbox{\hrule height 1.2pt\hbox{\vrule width 1.2pt\hskip 3pt
   \vbox{\vskip 6pt}\hskip 3pt\vrule width 0.6pt}\hrule height 0.6pt}\kern1pt}
\def\tdot#1{\mathord{\mathop{#1}\limits^{\kern2pt\ldots}}}

\def\pmb#1{\setbox0=\hbox{#1}%
  \kern-.025em\copy0\kern-\wd0
  \kern  .05em\copy0\kern-\wd0
  \kern-.025em\raise.0433em\box0 }

\catcode`@=11
\newcount\tagnumber\tagnumber=0

\immediate\newwrite\eqnfile
\newif\if@qnfile\@qnfilefalse
\def\write@qn#1{}
\def\writenew@qn#1{}
\def\w@rnwrite#1{\write@qn{#1}\message{#1}}
\def\@rrwrite#1{\write@qn{#1}\errmessage{#1}}

\def\taghead#1{\gdef\t@ghead{#1}\global\tagnumber=0}
\def\t@ghead{}

\expandafter\def\csname @qnnum-3\endcsname
  {{\t@ghead\advance\tagnumber by -3\relax\number\tagnumber}}
\expandafter\def\csname @qnnum-2\endcsname
  {{\t@ghead\advance\tagnumber by -2\relax\number\tagnumber}}
\expandafter\def\csname @qnnum-1\endcsname
  {{\t@ghead\advance\tagnumber by -1\relax\number\tagnumber}}
\expandafter\def\csname @qnnum0\endcsname
  {\t@ghead\number\tagnumber}
\expandafter\def\csname @qnnum+1\endcsname
  {{\t@ghead\advance\tagnumber by 1\relax\number\tagnumber}}
\expandafter\def\csname @qnnum+2\endcsname
  {{\t@ghead\advance\tagnumber by 2\relax\number\tagnumber}}
\expandafter\def\csname @qnnum+3\endcsname
  {{\t@ghead\advance\tagnumber by 3\relax\number\tagnumber}}

\def\equationfile{%
  \@qnfiletrue\immediate\openout\eqnfile=\jobname.eqn%
  \def\write@qn##1{\if@qnfile\immediate\write\eqnfile{##1}\fi}
  \def\writenew@qn##1{\if@qnfile\immediate\write\eqnfile
    {\noexpand\tag{##1} = (\t@ghead\number\tagnumber)}\fi}
}

\def\callall#1{\xdef#1##1{#1{\noexpand\call{##1}}}}
\def\call#1{\each@rg\callr@nge{#1}}

\def\each@rg#1#2{{\let\thecsname=#1\expandafter\first@rg#2,\end,}}
\def\first@rg#1,{\thecsname{#1}\apply@rg}
\def\apply@rg#1,{\ifx\end#1\let\next=\relax%
\else,\thecsname{#1}\let\next=\apply@rg\fi\next}

\def\callr@nge#1{\calldor@nge#1-\end-}
\def\callr@ngeat#1\end-{#1}
\def\calldor@nge#1-#2-{\ifx\end#2\@qneatspace#1 %
  \else\calll@@p{#1}{#2}\callr@ngeat\fi}
\def\calll@@p#1#2{\ifnum#1>#2{\@rrwrite{Equation range #1-#2\space is bad.}
\errhelp{If you call a series of equations by the notation M-N, then M and
N must be integers, and N must be greater than or equal to M.}}\else%
 {\count0=#1\count1=#2\advance\count1
by1\relax\expandafter\@qncall\the\count0,%
  \loop\advance\count0 by1\relax%
    \ifnum\count0<\count1,\expandafter\@qncall\the\count0,%
  \repeat}\fi}

\def\@qneatspace#1#2 {\@qncall#1#2,}
\def\@qncall#1,{\ifunc@lled{#1}{\def\next{#1}\ifx\next\empty\else
  \w@rnwrite{Equation number \noexpand\(>>#1<<) has not been defined yet.}
  >>#1<<\fi}\else\csname @qnnum#1\endcsname\fi}

\let\eqnono=\eqno
\def\eqno(#1){\tag#1}
\def\tag#1$${\eqnono(\displayt@g#1 )$$}

\def\aligntag#1\endaligntag
  $${\gdef\tag##1\\{&(##1 )\cr}\eqalignno{#1\\}$$
  \gdef\tag##1$${\eqnono(\displayt@g##1 )$$}}

\def\eqalignno#1{\displ@y \tabskip\centering
  \halign to\displaywidth{\hfil$\displaystyle{##}$\tabskip\z@skip
    &$\displaystyle{{}##}$\hfil\tabskip\centering
    &\llap{$\displayt@gpar##$}\tabskip\z@skip\crcr
    #1\crcr}}

\def\displayt@gpar(#1){(\displayt@g#1 )}

\def\displayt@g#1 {\rm\ifunc@lled{#1}\global\advance\tagnumber by1
        {\def\next{#1}\ifx\next\empty\else\expandafter
        \xdef\csname @qnnum#1\endcsname{\t@ghead\number\tagnumber}\fi}%
  \writenew@qn{#1}\t@ghead\number\tagnumber\else
        {\edef\next{\t@ghead\number\tagnumber}%
        \expandafter\ifx\csname @qnnum#1\endcsname\next\else
        \w@rnwrite{Equation \noexpand\tag{#1} is a duplicate number.}\fi}%
  \csname @qnnum#1\endcsname\fi}

\def\ifunc@lled#1{\expandafter\ifx\csname @qnnum#1\endcsname\relax}

\let\@qnend=\end\gdef\end{\if@qnfile
\immediate\write16{Equation numbers written on []\jobname.EQN.}\fi\@qnend}

\catcode`@=12

\catcode`@=11
\newcount\r@fcount \r@fcount=0
\newcount\r@fcurr
\immediate\newwrite\reffile
\newif\ifr@ffile\r@ffilefalse
\def\w@rnwrite#1{\ifr@ffile\immediate\write\reffile{#1}\fi\message{#1}}

\def\writer@f#1>>{}
\def\referencefile{%			  Stuff to write .REF file
  \r@ffiletrue\immediate\openout\reffile=\jobname.ref%
  \def\writer@f##1>>{\ifr@ffile\immediate\write\reffile%
    {\noexpand\refis{##1} = \csname r@fnum##1\endcsname = %
     \expandafter\expandafter\expandafter\strip@t\expandafter%
     \meaning\csname r@ftext\csname r@fnum##1\endcsname\endcsname}\fi}%
  \def\strip@t##1>>{}}

\def\citeall#1{\xdef#1##1{#1{\noexpand\cite{##1}}}}
\def\cite#1{\each@rg\citer@nge{#1}}	% Variable No. of args, separated by ","

\def\each@rg#1#2{{\let\thecsname=#1\expandafter\first@rg#2,\end,}}
\def\first@rg#1,{\thecsname{#1}\apply@rg}	% each@ag is a general purpose
\def\apply@rg#1,{\ifx\end#1\let\next=\relax%	  variable no. of arg. macro.
\else,\thecsname{#1}\let\next=\apply@rg\fi\next}% args separated by commas

\def\citer@nge#1{\citedor@nge#1-\end-}	% Check for M-N range (M and N numbers)
\def\citer@ngeat#1\end-{#1}
\def\citedor@nge#1-#2-{\ifx\end#2\r@featspace#1 % Single argument
  \else\citel@@p{#1}{#2}\citer@ngeat\fi}	% M-N range of arguments
\def\citel@@p#1#2{\ifnum#1>#2{\errmessage{Reference range #1-#2\space is bad.}%
    \errhelp{If you cite a series of references by the notation M-N, then M and
    N must be integers, and N must be greater than or equal to M.}}\else%
 {\count0=#1\count1=#2\advance\count1
by1\relax\expandafter\r@fcite\the\count0,%
  \loop\advance\count0 by1\relax%	  Loop from M to N
    \ifnum\count0<\count1,\expandafter\r@fcite\the\count0,%
  \repeat}\fi}

\def\r@featspace#1#2 {\r@fcite#1#2,}	% Eat spaces at beginning or end of arg
\def\r@fcite#1,{\ifuncit@d{#1}%		  Cite individual reference
    \newr@f{#1}%
    \expandafter\gdef\csname r@ftext\number\r@fcount\endcsname%
                     {\message{Reference #1 to be supplied.}%
                      \writer@f#1>>#1 to be supplied.\par}%
 \fi%
 \csname r@fnum#1\endcsname}
\def\ifuncit@d#1{\expandafter\ifx\csname r@fnum#1\endcsname\relax}%
\def\newr@f#1{\global\advance\r@fcount by1%
    \expandafter\xdef\csname r@fnum#1\endcsname{\number\r@fcount}}

\let\r@fis=\refis			% Save old \refis, redefine
\def\refis#1#2#3\par{\ifuncit@d{#1}%      Use two params #2 #3 to strip blank
   \newr@f{#1}%
   \w@rnwrite{Reference #1=\number\r@fcount\space is not cited up to now.}\fi%
  \expandafter\gdef\csname r@ftext\csname r@fnum#1\endcsname\endcsname%
  {\writer@f#1>>#2#3\par}}

\def\ignoreuncited{%   redefine \refis if ignoring uncited references
   \def\refis##1##2##3\par{\ifuncit@d{##1}%
     \else\expandafter\gdef\csname r@ftext\csname
r@fnum##1\endcsname\endcsname%
     {\writer@f##1>>##2##3\par}\fi}}

\def\r@ferr{\endreferences\errmessage{I was expecting to see
\noexpand\endreferences before now;  I have inserted it here.}}
\let\r@ferences=\references
\def\references{\r@ferences\def\endmode{\r@ferr\par\endgroup}}

\let\endr@ferences=\endreferences
\def\endreferences{\r@fcurr=0%		  Save old \endreferences, redefine
  {\loop\ifnum\r@fcurr<\r@fcount%	  Loop over refnum and produce text
    \advance\r@fcurr by 1\relax\expandafter\r@fis\expandafter{\number\r@fcurr}%
    \csname r@ftext\number\r@fcurr\endcsname%
  \repeat}\gdef\r@ferr{}\endr@ferences}

% Save old \endpaper, redefine it to write parting message.

\let\r@fend=\endpaper\gdef\endpaper{\ifr@ffile
\immediate\write16{Cross References written on []\jobname.REF.}\fi\r@fend}

\catcode`@=12

\citeall\refto		% These macros will generate citations
\citeall\ref		%
\citeall\Ref		%

\ignoreuncited
\rightline{NSF-ITP-91-01}
\vskip -8truept
\rightline{McGill/92-01}
\title{{\tif {}From polymers to quantum gravity:
triple-scaling in rectangular random matrix models
\phantom{s}}}
\author{\rm Robert C. Myers\footnote{${}^1$}{rcm@physics.mcgill.ca}}
\affil{Physics Department, McGill University
Ernest Rutherford Building
Montr\'eal, Qu\'ebec, CANADA H3A 2T8}
\author{\rm Vipul %
Periwal\footnote{${}^2$}{vipul@guinness.ias.edu. Also at Institute %
for Theoretical Physics, University of California, Santa Barbara, %
CA 93106-4030}}
\affil{Institute for Advanced Study
School of Natural Science
Princeton, New Jersey 08540-4920}
\abstract{Rectangular $N\times M$
matrix models can be solved in several qualitatively distinct
large $N$ limits,
since two independent parameters govern the size of the matrix.
Regarded as models of random surfaces, these matrix models
interpolate between branched polymer behaviour and two-dimensional
quantum gravity.
We solve such models in a `triple-scaling' regime in this paper, with
$N$ and $M$ becoming large independently.
A correspondence between phase transitions and
singularities of mappings from ${\bf R}^2$ to ${\bf R}^2$ is indicated.
At different critical points, the
scaling behavior is determined by: i) two decoupled ordinary differential
equations; ii) an ordinary differential equation and a finite
difference equation; or iii) two coupled partial differential equations.
The Painlev\'e II equation arises (in conjunction with
a difference equation) at a point associated with branched
polymers.  For critical points described by partial differential
equations, there are dual
weak-coupling/strong-coupling
expansions.  It is conjectured that the new physics is related to
microscopic topology fluctuations.}

\endtopmatter
\baselineskip=15truept

\centerline{Introduction}
\bigskip

Advances have been made in the study of noncritical string
theory with the discovery of
the double-scaling limit\cit{big3}
in random matrix models. The double-scaling
limit correlates the approach of the
couplings in the matrix model to certain critical values with the
large-$N$ limit.
We give herein the solution of rectangular $N\times
M$ matrix models in a `triple-scaling' regime---$N$ and $M$ are taken to
infinity independently in correlation with the
approach to critical matrix couplings.
The surface interpretation of these models in certain limits has
been described in previous work\cit{arley1,arley2}.
%---some further
%observations will be given at appropriate junctures in the following.
The interest in these models is two-fold: (a) they interpolate between
the rather different physics of random surfaces, and branched polymers,
and (b) they exhibit critical behaviour
%described by partial differential equations
involving two scaling variables.
%While we have found several interesting `phenomena' in the
%solutions of various matrix models, a unified understanding is still
%missing.
We list explicitly
the novel results/techniques contained in the present work:
\item{1.} We find phase transitions in correspondence to singularities
of mappings from ${\bf R}^2$ to ${\bf R}^2.$
\item{2.} We use recursion relations relating polynomials for different
measures on the half-line.  The significance of this lies in the fact
that the derivative terms in the differential equations that describe
critical behaviour can arise from either recursion relations relating
different values of $N$ or $M$ (or linear combinations of these two
parameters).
\item{3.} The resulting scaling behavior may be governed
by: (i) two decoupled ordinary differential
equations; (ii) an ordinary differential equation and a finite
difference equation; or (iii) two coupled partial differential equations.
\item{4.} In case (iii) above,
the equations admit weak-coupling solutions in
either variable, which are simultaneously strong-coupling expansions in
the other variable.

Since the results and techniques
given in the present paper are largely unrelated to the contents of
Ref.'s~\cite{arley1,arley2}, it can be read independently of
our previous work.  To put it in context, we mention that
in work with Anderson\cit{arley1,arley2}, we solved rectangular $N\times
M$ matrix models in two distinct {\it double}-scaling
limits, with either
$M$ (limit I)\footnote{${}^\dagger$}{This case was also investigated in %
Ref.'s \cite{yoneya,div} with $M=1$.} or $N-M$
(limit II)\footnote{${}^\ddagger$}{Ref.'s~\cite{tim,div} analyzed %
this case with $N\!-\!M=0$.} held fixed as $N$ diverges.
The limit I models describe dynamically
branching polymers\cit{yoneya,arley2}, while the limit II models yield
two-dimensional random surfaces\cit{tim,arley1}.
In Ref.~\cite{rvec}, one of us (R.C.M.) solved coupled vector models
in a triple-scaling limit, which has some features in common with
the rectangular matrices solved in the present work.
%However, the rectangular matrix models appear very restricted,
%in comparison,
%in the types of critical behaviour that can occur.
We wish to draw attention to the work of G.M. Cicuta and
collaborators\cit{bell},
who solved rectangular matrix models in the planar limit several years
ago\footnote{${}^*$}{We regret that we were unaware of this work when %
Ref.'s \cite{arley1,arley2} were published.}, and to the work of
J. Minahan, who has
solved matrix models for surfaces with boundaries\cit{joe1,joe2joe3}---some
of the equations we shall obtain in the following also arise in his studies
(see sect. 4).

The organization of this paper is as follows:
\item{\bb} Sect.~1 contains some notation, the Jacobian of interest, and
a list of properties of the orthogonal polynomials used. We also present
identities involving the recursion coefficients, necessary to solve the
triple-scaling limit.
\item{\bb} In sect. 2, we give the planar limit of the string equations.
We discuss the identification of the critical models with singularities
in two-dimensional mappings. We also give a brief discussion of the
saddle point analysis, which provides a complementary description
of the planar limit.
\item{\bb} Triple-scaling {\it ans\"atze} are introduced in
sect.~3. We illustrate the construction of scaling solutions with
an explicit example. For this example, the associated two-dimensional
mapping can be decomposed into two separate one-dimensional singularities,
and the resulting scaling solution involves two decoupled
ordinary differential equations in two independent variables.
\item{\bb} In sect.~4, we provide examples of new scaling equations, whose
general form is a single ordinary differential equation which includes
the second scaling variable as a parameter. The scaling functions
satisfy finite difference equations for shifts in this second variable.
Two of these examples are related to the simplest intrinsically
two-dimensional singularity, the Whitney pleat\cit{sade}.
\item{\bb} Sect.~5 displays a new family of multicritical models for which
i) the associated singularities are again intrinsically two-dimensional,
ii) the scaling equations are two coupled partial differential equations in
the two scaling variables, and
iii) the equations admit two qualitatively different weak-coupling
expansions.
\item{\bb} Some concluding remarks and conjectures are given in sect.~6.

\section{1. Notation and polynomial recursion relations}

\def\SO(#1){{\rm SO}_{#1}}
\def\Sp(#1){{\rm Sp}_{#1}}
\def\U(#1){{\rm U}_{#1}}
We define first matrices of the following form:
$$H\equiv \left(\matrix{0&T\cr T^\dagger&0\cr}\right),\eqno(def)$$
where $T$ is an $N\times M$
matrix with complex entries.
{In the following, we will assume that $N\ge M$}
without any loss of generality.
These matrices have a natural action of
U$(N)\times$U$(M),$ which can be used to bring $T$ to the form
$$T=\left(\matrix{X_{\ssc M}\cr 0_{\ssc (N-M)\times M}\cr}\right),
\eqno(can)$$
where $X_{\ssc M}$ is a real diagonal $M\times M$ matrix ${\rm
diag}(x_{\ssc 1},x_{\ssc 2},\dots,x_{\ssc M}).$
Analogous statements can be made about matrices with
real or quaternionic entries, as well as circular ensembles obtained
by exponentiating such matrices\cit{arley1}.
All the steps in the following can be repeated for these ensembles
as well, albeit with some technical complications\cit{symp1,bren,sump}.

\def\cz{{\cal Z}}
\def\dif{{\rm d}}

We wish to study the free energy of the partition function defined as
$$\cz\equiv\int \dif H \exp(-\be \tr V(H^2)),$$
where $\dif H$ is the Haar measure, and the matrix potential is a polynomial
of the form:
$V(H^2)=\sum_{p=1}^L{a_p\over2p}\,H^{2p}$.
(Note that $\tr H^{2k+1}=0$.) Such matrix integrals
lead to theories of random surfaces, via Feynman diagram
expansions\cit{thooft}.
The invariance of $\cz$ under U$(N)\times$U$(M)$ leads to
the following integral:
$$\cz = \int_{-\infty}^\infty\prod_{i=1}^{M} \left[\dif x_i\ \exp(-2\beta
V(x_i^2))\,\right]\ \ \prod_{i=1}^{M}|x_i|^{2(N-M)+1}\prod_{1\le i < j \le M}
\left(x_i^2-x_j^2\right)^2\eqno(part)$$
where overall factors, which do not affect the critical behavior, have been
dropped. {}From the Jacobian appearing in \(part), it is clear that the
partition function
is a function of $P\equiv N-M,$ and $M,$ independently.
%In limit I, $P$
%is finite as $N$ goes to infinity, and in limit II, $M$ is finite.
\def\P#1#2{P^{(#1)}_{#2}\!(y)}
\def\h#1#2{h^{(#1)}_{#2}}
\def\rs#1#2{S^{(#1)}_{#2}}
\def\ra#1#2{\al^{(#1)}_{#2}}
\def\rt#1#2{\theta^{(#1)}_{#2}}
\def\rp#1#2{\phi^{(#1)}_{#2}}

An obvious change of variables from $x$ to $y=x^2$ simplifies the analysis
of \(part). It is then clear that we need orthogonal polynomials on a
half-line, and relations between polynomials for measures related
by the insertion of powers of $y.$  Some of these are derived in
Chihara\cit{chihara}.  Let
$$\dif \mu[l] \equiv y^l \exp\left(-2\beta V(y)\right),$$
where one assumes the measure has support in
$[0,\infty).$\footnote{$^\dagger$}{The measure will be divergent at
the latter end-point for several critical models, in which case the following
identities must be regarded as analytically continued from the
convergent cases.}
Denote orthogonal polynomials for such a measure by $\P ln,$
such that
$$\int_{0}^{\infty}\!\dif \mu[l]\ \P l n\ \P l m = \delta_{n,m}\, \h l m\ ,$$
where $\P ln\equiv y^n+$ lower order terms.
Multiplying by $y$ yields the usual three-term recursion relation
$$y\,\P l n =\P l{n+1} + \rs l n \P l n + \ra l n\P l{n-1} ,\eqno(cur1)$$
where $\ra l n =\h l n /\h l{n-1}$. There are also simple recursion
relations relating polynomials defined for measures with different
values of $l$
$$\eqalign{\P l n &=\P {l+1}n +\rp l n\, \P {l+1}{n-1} ,\cr
y\,\P {l+1}n &=\P l{n+1} + \rt l{n+1}\, \P l n, \cr}\eqno(cur2)$$
where
$$\rp l n ={\h l n\over\h{l+1}{n-1}}\qquad{\rm and}\qquad
\rt l{n+1} ={\h {l+1}n\over\h l n}=-{P^{(l)}_{n+1}\!(0)\over P^{(l)}_n\!(0)}
\ .\eqno(deaf)$$
Our scaling analysis in the following will be made in terms of the
recursion coefficients, $\rp ln$ and $\rt ln$.
The coefficients appearing in \(cur1) are easily determined by
$$\rs l n = \rt l{n+1} +\rp l n \qquad{\rm and}\qquad \ra l n =\rt l n\rp l n
\ \ .\eqno(old)$$
One can show that $\rt ln$ and $\rp ln$ appear as recursion
coefficients in the formula analogous to \(cur1) for orthogonal polynomials
defined for $x$, rather than $y$.

Further useful relations between the coefficients are
$$\eqalignno{
\rp l n -\rp{l+1}{n-1} &=\rt {l+1}n -\rt l n ,&(t1)\cr
\rp {l-1}n\,\rt {l-1}{n+1} &=\rp l n\,\rt l n\ \  .&(t2)\cr}$$
These relations are independent of the matrix potential, $V(y)$,
and their derivation is elementary, so we shall not give it here.
One can think of \(t1,t2)\ are discrete analogues of
`zero-curvature' conditions (\ie they ensure that the
operations of shifting the indices $l$ and $n$ commute).
Their importance in the solution of these models is quite considerable.
Indeed in section 5, we shall find a KdV-like partial
differential originating from the study of these relations.  To
this equation are associated an infinite set of triple-scaling
multicritical points. These identities become trivial in the planar
limit, with $l=P$, $n=M$, and
$\be\rightarrow\infty$ keeping $P/\be$ and $M/\be$ finite (see sect.~3).

Using the properties listed above, one can derive two independent
`string equations':
$$\eqalignno{
{2n+l+1\over 2\be}\h ln &= \int\!\dif \mu[l]\ y\, V'(y)\ \P ln\,\P ln
&(dyn1)\cr
{n\over 2\be}\h{l+1}{n-1} &= \int\!\dif \mu[l]\  V'(y)\left(\P l{n}^2 +
\rt l{n} \P l{n}\P l{n-1}\right)\ .&(dyn2)\cr}$$
The independence of these equations is obvious since $l$ and $n$ are
completely independent parameters.
Using the operator techniques introduced in Ref.~\cite{gross3}, one can
reexpress these identities as
$$\eqalignno{
&{2n+l+1\over 2\be} = \int_0^{2\pi} {\dif\la\over2\pi}\ \bar{y}(n-i\part_\la)
\, V'[\bar{y}(n-i\part_\la)]\cdot 1_l&(dyn1b)\cr
&{n\over 2\be} = \rp l{n} \left(\int_0^{2\pi}
{\dif\la\over2\pi}\ V'[\bar{y}(n-i\part_\la)]
\cdot 1_l\ +\ \rt ln \int_0^{2\pi} {\dif\la\over2\pi}
\ e^{-i\la}V'[\bar{y}(n-1-i\part_\la)]
\cdot 1_l\right)&(dyn2b)\cr}$$
where
$$\bar{y}(n-i\part_\la)=e^{i\la}+\hat{\th}(1+n-i\part_\la)
+\hat{\phi}(n-i\part_\la)+e^{-i\la}\hat{\th}(n-i\part_\la)
\hat{\phi}(n-i\part_\la)\ \ .$$
The subscript $l$ on $1_l$ indicates that all of the
recursion coefficients carry the same superscript $(l)$
(\ie $\hat{\th}(n-i\part_\la)\cdot 1_l =\rt ln$). Note then
that these identities only involve coefficients with differing
subscripts.
These equations are also non-trivial in the planar limit. Both
of these features contrast them from the potential independent identities,
\(t1) and \(t2).
%The identities \(t1) and \(t2) also provides an equation which
%is {\it trivial}
%in the planar limit, is not integrable at the discrete level, but
%provides a relation between the scaling functions associated
%with $S,\al$ in the scaling limit.

\section{2. Planar limit}

We begin by considering the planar limit, in which only the contributions
of surfaces with spherical topology survive. Explicitly in the recursion
coefficient identities above,  we set
$l=P$ and $n=M$, and consider the limit
$\be\rightarrow\infty$ keeping $g\equiv N/\be$ and $q\equiv M/\be$ finite.
We assume that the recursion coefficients have a smooth limit,
$\rt {\ssc P\pm i}{\ssc M\pm j}\raro\th$
and $\rp {\ssc P\pm i}{\ssc M\pm j}\raro\phi$. Clearly
in this case,  eq.'s \(t1,t2) vanish trivially.
On the other hand, eq.'s \(dyn1b,dyn2b)
become algebraic, and can be interpreted as giving the extrema of the
following functional
$$W(\th,\phi)=U(\th,\phi)-g\,\ln\th-q\,\ln\phi$$
where
$$U(\th,\phi)=2\int_0^{2\pi} {\dif\la\over2\pi}\ V(e^{i\la}+\th
+\phi+e^{-i\la}\th\phi)\ \ .$$
Explicitly, given a potential $V(y)=\sum_{p=1}^L{a_p} y^p/(2p)$, one finds
$$ U(\th,\phi) =\sum_{p=1}^L{a_p\over p}\ \sum_{k=0}^{p}
{p\choose k}^2\th^k\phi^{p-k}\ \ .\eqno(whoopee)$$
Recalling the relations \(old) of $\th$ and $\phi$ to the usual recursion
coefficients in \(cur1), note the similarity of $U$ to the
functional $\Omega$ introduced in Ref.~\cite{gross3}.

The critical points can be identified as follows:
The planar equations take the form
$$g=\th\,\part_\th U(\th,\phi)\qquad {\rm\ and\ }\qquad q=\phi\,\part_\phi
U(\th,\phi)\ \ .\eqno(plane)$$
In the case of hermitian matrix models\cit{gross3},
one finds a single equation
$g=g(\al)$. In that case, the critical values of the coupling constants
are chosen such that ${\dif^i g\over\dif\al^i}=0$ for $i=0,\ldots,k$.
Thus the critical points are identified as points where $\al=\al(g)$ is
nonanalytic, or alternatively as points where the map $g\mapsto\al$
is singular. Given this point of view, it is natural to identify the
the critical points in the present case as points where the two-dimensional
map $(g,q)\mapsto(\th,\phi) $ becomes singular.
Such points occur at the vanishing of the Jacobian
determinant\cit{sade}:
$$0=|J|=\left|\matrix{{\part g\over\part\th}&{\part g\over\part\phi}\cr
        {\part q\over\part \th}&{\part q\over\part\phi}\cr}
            \right|\ .\eqno(jake)$$
Multicritical behavior is produced by demanding the vanishing of various
higher derivatives of $g$ and $q$. In terms of the functional $W$, these
(multi)critical points correspond to degenerate extrema. In the scaling limit
which will be of interest below, one does
not focus on the degenerate point exactly. Rather
the singularity is resolved by also
perturbing $g$ and $q$ near their critical values\cit{big3}.
Schematically one might find
$$W\approx\sum A_i (\hat{\phi}_c-\hat{\phi})^{p_i}
(\hat{\th}_c-\hat{\th})^{k_i} - (\hat{g}_c-\hat{g})
(\hat{\th}_c-\hat{\th})-(\hat{q}_c-\hat{q})
(\hat{\phi}_c-\hat{\phi})\eqno(singsong)$$
where $p_i$ and $k_i$ are integers, and $A_i$, arbitrary
constants. The variables in \(singsong)
are denoted with $\hat{\ }$ to indicate that they can be arbitrary
{\it linear} combinations of $\phi$, $\th$, $g$ and $q$.
Generically the singularity
will most naturally be parameterized in terms of various linear combinations,
rather than $\phi$, $\th$, $g$ and $q$ directly.
This analysis of the critical points is identical to that for coupled
vector models\cit{rvec}. In that case, the functional $W$ is
replaced by an effective potential $\widehat{V}$, which is of direct
physical relevance in those models. The coefficients of the
polynomial terms in $\widehat{V}$ are all completely independent. In contrast,
\(whoopee) shows that many of these coefficients in $W$ are directly related.
%Practical experience with calculations for the rectangular matrix models,
%has shown us that
This restricts the possible singularities that arise.
The essential two-dimensional character of the critical points
leads to a much richer structure than appears for ensembles of square
matrices, though. We do not have a complete classification of
all possible singularities, but in the following, we will provide
various explicit examples which illustrate
very distinct scaling behaviors. Each of
these examples extend to an infinite family of higher multicritical
models, some of which will be discussed further
in Ref.~\cite{lafrance}.

Before proceeding with examples of triple-scaling behavior, we would like to
briefly discuss the saddle point analysis, because it provides a
more physically intuitive description of the planar limit, and
of the critical behavior.
First, we mention that using the saddle point approximation,
the planar limit of rectangular matrix models
with non-vanishing $a_1$ and $a_2$ was completely solved in Ref.~\cit{bell}.
The following discussion applies for general potentials.
\def\print{-\hskip-1.1em\relax\int}

The partition function can be written as
$$\cz_{\ssc N,M} = \int_0^\infty\prod_{i=1}^{M}
\left(\dif y_i^{}\,y_i^{N-M}\, \exp\left[-2\beta V(y_i)\right]\,\right)\
\prod_{1\le i < j \le M} \left(y_i-y_j\right)^2\eqno(part1)$$
with $V(y)=\sum_{p=1}^L\!{a_p} y^p/(2p)$.
The entire integrand can be expressed as $\exp(-\beta^2 E)$, where
$$\beta E(\{y_i\}) = 2\sum_{i=1}^M V(y_i) - {N-M\over\beta}\sum_{i=1}^M
\ln y_i-{1\over\beta}\sum_{i\not=j;i,j=1}^M \ln|y_i-y_j|\ \ .$$
The last term above arises from
the Vandermonde factor in the Jacobian,
and  can be interpreted as a repulsive electrostatic potential
between different particles at $y_i$ in a one-dimensional system.
The second term also coming from the Jacobian,
represents a further electrostatic repulsion away
from the origin, whose strength is controlled by $P=N-M$. This repulsion
combines
with the external potential $V$ to balance the Coulomb repulsion of the
particles, and confine them to some interval $[a,b]$ at equilibrium.
We will assume $0\le a\le b$.\footnote{$^\dagger$}{The following analysis %
is only correct for $a=0$ if the eigenvalue density vanishes there. A proper %
treatment for a nonvanishing density at the origin is most easily done in %
terms of the original variables $x$ rather than $y$\cit{tim2,arley2}.}
Evaluating
\(part1) for large $\beta$ in the saddle
point approximation, requires finding a
configuration of `eigenvalues' satisfying
$${\partial V(y_i)\over {\partial y_i}} - {N-M\over {2\beta y_i}} -
{1\over \be}\sum_{j\not=i;j=1}^M {1\over y_i-y_j}= 0\ \ .\eqno(sad)$$
Defining a density of
eigenvalues, $\rho(y),$ in the usual manner\cit{bipz}, one finds
$$ V'(y) - {g-q\over 2y} = q\print_{a}^{b} {1\over y-z}
\rho(z)\dif z.\eqno(move)$$
%where we have reintroduced $q=M/\be$ and $g =N/\be.$
Solving this equation for $\rho(y)$ is an example of the Hilbert problem
on an arc\cit{integral}.
The results can be expressed as constraints
on the function
$$F(z)={1\over q}\left(V'(z)-{g-q\over2z}-u(z)\sqrt{(z-b)(z-a)}\right)$$
where $u(z)=\sum_{k=-1}^{L-2}h_k z^k$. The constants $h_k$, $a$ and $b$
are fixed by demanding that $F(z)\raro1/z$ as $|z|\raro\infty$,
and that $F(z)$ remains finite as $|z|\raro 0$.
The eigenvalue density on $[a,b]$ is then given by
$$\rho(y)={1\over\pi q}u(y) \sqrt{(b-y)(y-a)}\ \ .\eqno(thickasbrick)$$
By examining the equations fixing $a$ and $b$, one finds that they
are identical to the planar string equations \(plane) when one makes the
identification
$$ \th=\quart(\sqrt{b}+\sqrt{a})^2\qquad{\rm and}\qquad
\phi=\quart(\sqrt{b}-\sqrt{a})^2\ \ .\eqno(bound)$$
Further using this identification, the coefficients $h_k$ can be written as
$$h_k={1\over 2}\sum_{p=0}^{L-2-k}a_{p+k+2}\ \sum_{l=0}^{p}{p\choose l}^2
\th^{l}\phi^{p-l} \eqno(reminis)$$
which is a form similar to that for $U(\th,\phi)$ in \(whoopee).

The aim is to evaluate the planar free energy,
$$E_0 = q\int_a^b \!\dif y\,\rho(y)\,\left[ 2V(y)-(g-q)\ln|y|\,\right] -
q^2\int_a^b\!\dif z\,\dif y\
\rho(z)\,\rho(y)\ {\rm ln} |z-y|\ \ .\eqno(hotplan)$$
Integrating \(move), one observes that
$$q\int_a^b\!\dif z\ \rho(z)\, {\rm ln}|y-z| = V(y)
- {g-q\over 2}\ln|y| - K,$$
where one can set
$$K = V(a)- {g-q\over 2}\ln|a| - q\int_a^b\!\dif z\, \rho(z)\,{\rm ln}|z-a|
\ \ .$$
Then \
$$E_0 = q \int_a^b\!\dif y\, \rho(y) \left[ V(y) -{g-q\over2}\ln|y|- q\,
\ln|y-a|\right] + qV(a)-  {g-q\over 2}q\,\ln|a|\ .$$

%{}From this point of view, the
Critical behaviour corresponds to
a nonanalytic dependence of the free energy $E_0$ on the coupling
constants in $V$ or $q$. For fixed couplings, this can be expressed
as nonanalyticity in $g$ and $q.$  Roughly we want
$$E_{sing} \sim (\hat{g}_c-\hat{g})^\gamma(\hat{q}_c-\hat{q})^\delta
\ \ .$$
Such nonanalytic behavior arises at the melting phase transition,
where the eigenvalue density \(thickasbrick) acquires zeroes
at the boundaries, $a$ and/or $b$,
beyond those evident in \(thickasbrick). Using eq.'s \(whoopee,plane,
bound,reminis), this mechanistic description is related to \(jake)
%one
%can make the connection of this discussion with the previous analysis of
%the critical points
by observing that $|J|=4a\,b\,u(a)\,u(b)$. %Thus the
%critical models identified in \(jake) with $|J|=0$
%are precisely those for which these extra
%zeroes appear in \(thickasbrick) since $u(a)$ and/or $u(b)$ vanish.

\def\part{\partial}
\def\hnu{\hat{\nu}}

\def\var{\Delta}
\section{3. Triple-scaling and Split Singularities}

In this section,
we discuss triple-scaling for the case of split singularities.
We use the latter term to describe
two-dimensional singularities which can be separated into two independent
one-dimensional singularities.
In double-scaling physics\cit{big3}, many relations between quantities
associated with critical behaviour follow from the basic {\it ansatz}
that
$${N\over \be} = \left[{N\over \be}\right]_{\!c}
- \be^{\nu - 1} t,\eqno(accord)$$
where $t$ is the double-scaling variable associated with the continuum
theory.  So long as $\nu>0,$ it is reasonable to assume that the
difference equations that characterize the matrix model at finite $N$
are transmuted into differential equations in the variable $t$ as
$\be\rightarrow\infty.$
In triple-scaling physics, as in Ref.~\cite{rvec},
we have two variables in the planar limit
that can play the r\^ole of $N,$ and thus \(accord) is naturally
extended to
$${X\over \be} = \left[{X\over \be}\right]_{\!c} - \be^{\nu - 1} t
\qquad{\rm and}\qquad
{Y\over \be} = \left[{Y\over \be}\right]_{\!c} - \be^{\hnu - 1} s
\eqno(cordee)$$
where $X$ and $Y$ are linearly independent combinations of $M$ and $N$
determined for a given singularity.

In our first example,
rather than imposing simply $|J|=0$ as in \(jake),
we require the stronger condition that the individual derivatives
in the Jacobian matrix all vanish (\ie ${\part g\over\part\th}=0=
{\part g\over\part\phi}={\part q\over\part\th}={\part q\over\part\phi}$
).\footnote{$^\dagger$}{Critical models with the simplest %
tuning to produce $|J|=0$ have more involved scaling solutions---see %
the following sections.}
Thus the expansion of the planar equations \(plane) about the critical
point will begin at second order in both variations: $\var\phi=\phi-\phi_c$
and $\var\th=\th-\th_c$.
%Note that actually
One finds that when any two of the
derivatives are set to vanish, the remaining two vanish automatically.
Tuning for this critical point, therefore, imposes fewer
constraints on the matrix potential than one might expect.
The minimal potential only requires that the first three coefficients
be nonvanishing.

%First observe that
There is enough freedom in the planar
equations to rescale the variables at any critical point to set
$g_c=\left[{N/\beta}\right]_c=1$
and $\th_c=1.$ %and for simplicity then
We make this
choice. We also set $\phi_c=y^2$
%to emphasize that it should be a
since it is non-negative---see
%quantity as is apparent from
either \(bound) or \(deaf).
With these choices, the matrix potential
can be tuned so that the desired derivatives vanish
by setting:
$$a_1=-3{1+y^4\over3y^2-1}\ ;\qquad a_2=3{1+y^2\over3y^2-1}\ ;
\qquad a_3=-{1\over3y^2-1}\ \ .$$
One finds that $q_c=\left[M/\beta\right]_c=y^4(3-y^2)/(3y^2-1)$.
%Therefore so that
%while the critical value of
$\phi_c$ remains free---%unfixed,
different values
correspond to approaching the large-$N$ limit with rectangular matrices of
different proportions. %Requiring that
$g_c=1\ge q_c\ge 0$ implies %, imposes the restriction
that $3\ge y^2\ge 1$. Thus %Note that in this range,
the critical potential is unbounded from below.
Expanding \(plane) about the critical point, %one finds
to leading order
$$\eqalign{
\var g&= -{3\over3y^2-1}(\var\phi^2+2\var\phi\var\th+y^2\var\th^2)\cr
\var q&=-{3\over3y^2-1}(\var\phi^2+2y^2\var\phi\var\th
+y^2\var\th^2)\cr}\eqno(quads)$$
where $\var g=g-g_c$, $\var q=q-q_c.$
Eq. \(quads) can be rewritten as
$$y \var g\pm \var q=-{3\over3y^2-1}(y\pm1)(y\var\th\pm\var\phi)^2\ ,
\eqno(rerite)$$
so %that it is actually
one has two independent quadratic singularities.

Eq. \(rerite) also shows that the natural parameters
in which to construct the triple scaling solution are $y g\pm q$
and $y\th\pm\phi$. First for \(cordee), we define $t,s$ by
$$\eqalign{
{y N+M\over \be} &= \left[{y N+M\over \be}\right]_{\!c} - \be^{\nu - 1} t\cr
{y N-M\over \be} &= \left[{y N-M\over \be}\right]_{\!c} -
 \be^{\nu - 1} s\ \ .\cr}
\eqno(cordee2)$$
The same exponents are chosen in both expressions,
since the independent singularities are
the same. With these {\it ans\"atze}, one can Taylor expand the
differences in $M$ and $P$ that
arise in eq.'s \(t1,t2,dyn1b,dyn2b) using
$$
\part_M=-({y+1})\del\part_t-({y-1})\del\part_s
\qquad{\rm and}\qquad
\part_P=-{y}\del\part_t-{y}\del\part_s
\eqno(derivative)$$
where $\del=\be^{-\nu}$.
The natural scaling {\it ans\"atze} for the recursion coefficients are then
$$\eqalign{
y\rt{P+l}{M+n}+\rp{P+l}{M+n}&=y(1+y)-\exp[-\Delta_{l,n}]
\sum_{q=2}\del^q h_q(t,s)\cr
y\rt{P+l}{M+n}-\rp{P+l}{M+n}&=y(1-y)-\exp[-\Delta_{l,n}]
\sum_{q=2}\del^q k_q(t,s)\ .\cr}
\eqno(solve)$$
where $\Delta_{l,n}\equiv\del\left[
{n}((y+1)\part_t+(y-1)\part_s)+l\,{y}(\part_t+\part_s)\right].$

We find that $\nu={1\over 5}$.
The potential independent equations \(t1,t2) yield:
$$\part_sh_2(t,s)=0=\part_tk_2(t,s)\ .\eqno(splinter)$$
Then %the remaining equations
\(dyn1b,dyn2b) give %lead to:
$$\eqalign{
s&=3{y-1\over3y^2-1}k_2(s)^2-{y^2(y-1)^2\over3y^2-1}\part_s^2k_2(s),\cr
t&=3{y+1\over3y^2-1}h_2(t)^2-{y^2(y+1)^2\over3y^2-1}\part_t^2h_2(t)\ .\cr}$$
This result, that the scaling solution is governed by two independent
Painleve I equations, may have been anticipated given that the critical
point consists of two separate quadratic singularities. Such
one-dimensional singularities
lead to the Painleve I equation for hermitian matrix models\cit{big3}.
{\it A priori} though, there is no reason that the derivative terms
in the differential scaling equations should respect this splitting of the
singularity. The crucial result implying %in achieving
this separation is \(splinter), which is solved by
$k_2=k_2(s)$ and $h_2=h_2(t)$.

%The above results must still be related to
What is the free energy of the matrix
model? % Using the fact that
$\cz_{\ssc M+1,P}/\cz_{\ssc M,P}=(M+1)\h PM$
and \(deaf) %the definitions of the recursion coefficients in terms of the
imply that %orthogonal polynomial weights, one finds that
$$\eqalign{
\ra PM&\approx{\cz_{\ssc M+1,P}\cz_{\ssc M-1,P}\over\cz_{\ssc M,P}^2}
\approx\exp(-\part^2_M F)\cr
\rt P{M+1}&\approx{\cz_{\ssc M+1,P+1}\cz_{\ssc M,P}\over\cz_{\ssc M,P+1}
\cz_{\ssc M+1,P}}\approx\exp(-\part_M\part_P F)\ .\cr}
\eqno(freebird)$$
Combining these results with \(derivative) and \(solve), one is able to show:
First,
$\part_t\part_sF$ is a constant. Then up to some possible analytic terms,
one finds
$$\part^2_sF={k_2(s)\over2 y^2(y-1)}\qquad{\rm and}\qquad
\part^2_tF={h_2(t)\over2 y^2(y+1)}\ .$$
Thus the nonanalytic contributions to the free energy are divided into
two completely separate pieces, both with asymptotic
expansions identical to that found for ordinary two-dimensional
gravity\cit{big3}. The relevant expansion parameters in the present
case are $t^{-{5\over 2}},s^{-{5\over2}}=\beta^2\left({yN\pm M\over\beta}-
\left[{yN\pm M\over\beta}\right]_c\right)^{-{5\over2}}$.
%This brings to mind the doubling of the
%scaling functions that arises in hermitian matrix models\cit{double}.
%Unlike the latter though, we emphasize that
The appearance of two scaling functions should not be confused
with the doubling found in hermitian
matrix models\cit{double}, since in the present case they are
functions of completely independent parameters.
%The equations in the present case involve two completely
%independent scaling parameters.
Furthermore, both scaling equations are
inhomogenous in the scaling variables without the appearance of any
nonuniversal coefficients\cit{simon}.
%In fact, we note that
By rescaling
$h_2$, $k_2$, $t$ and $s$ appropriately,
the results for all different values of $\phi_c$
can be reduced to
$$\eqalign{
s&=k_2(s)^2-{1\over6}\part_s^2k_2(s)\ ;\qquad
\part^2_sF=k_2(s)\cr
t&=h_2(t)^2-{1\over6}\part_t^2h_2(t)\ ;\qquad
\part_t^2F=h_2(t)\ .\cr}$$
Therefore all of these models with matrices of different proportions
produce identical continuum physics.
%(These results are also unaffected by introducing higher order coefficients
%in the matrix potential while maintaining the vanishing of derivatives in
%the Jacobian matrix).

The natural generalization for these split singularities would be to
tune
the potential to produce separate one-dimensional singularities of
independent orders. However, such singularities do not arise
%seem to be relevant for
in the triple-scaling solution of rectangular matrix models.
The example given
here does generalize to a family of split singularities of the same order,
$$y\var g\pm\var q\propto(y\var\th\pm\var\phi)^n\eqno(splitn)$$
where $n$ is a positive integer.
Implicitly we include the `topological' model with $n=1$, but of course
this does not correspond to a singular mapping.
The only assumption involved in producing \(splinter) was that the scaling
parameters appear with the same exponents. Since those results were
derived from the potential independent equations \(t1,t2), they will
also apply for the higher singularities considered in \(splitn).
Hence one will recover $k_2=k_2(s)$ and $h_2=h_2(t)$ for all of these critical
points, and the scaling behaviour will then be governed
by two independent ordinary differential equations
of the form appearing in double-scaled matrix models.
Note that at least up to $n=4$, we found that tuning to increase the
power of the singularity in \(splitn) by one, only requires introducing two
extra coefficients in the minimal potential. This fact is an example
of how the interrelatedness of the
coefficients in the functional $U(\th,\phi)$ in \(whoopee) restricts
the set of singularities that can be attained
in these matrix models.

At the beginning, we found that $g_c>q_c>0$ required
$3>y^2>1$. One may ask what happens at the end
points of this interval. With $y^2=3$, the previous analysis and solution
remain unchanged. It is only an exceptional point since $q_c=[M/\beta]_c=0$.
%One should not mistake this as meaning
This does not mean that the number of eigenvalues
is zero or even finite. {}From \(cordee), one finds that
$M=\beta^{{1\over 5}}(s-t)/2$. Therefore the number of eigenvalues
diverges, but at a much slower rate than either $N$ or $\beta$. At $y^2=1$,
the previous analysis fails. One sees in \(rerite) that one of
$y\var g\pm\var q$ vanishes to the given order in the expansion of planar
equations. Hence the nature of the singularity changes at this
point, and higher order terms must be included in the expansion
\(quads) for a proper
analysis. In this case, the scaling solution is similar to that presented
in the section 5, as will be discussed in Appendix A.

%One may wish to
Consider the saddle point solution for the critical point
solved here. Using \(thickasbrick) and \(reminis),
the eigenvalue density has two extra zeroes, one at each of the two
boundaries (\ie $u(a)=0=u(b)$). %, as might be expected. Unfortunately
The saddle point
analysis leading to this result is {\it not} valid in the present case as
is evident from \(bound), which requires $\theta_c\ge\phi_c$.
As discussed above, the scaling analysis is valid for $\phi_c=y^2>1=\th_c$.
%This indicates that
Therefore the triple scaling solution presented here corresponds
to a more involved analytic continuation of the matrix model than considered
for the saddle point analysis presented in the previous section.
%Note though that
This complication does not hold for all of the split
singularity critical points. One finds that $\theta_c\ge\phi_c$ for the
odd powers in \(splitn).

\section{4. Parametric Solutions}

In this section we exhibit some novel critical phenomena, associated with
intrinsically two-dimensional singularities. We begin with a general
analysis of the planar limit to identify critical points
for the simplest nontrivial potential,
$V(y)={a_1}\,y/2+{a_2}\,y^2/4$. We focus on the scaling solution
of two higher order multicritical points. %These critical
%points are exceptional in the same sense that $y^2=1$ was an exceptional
%point in the previous section.
The scaling analysis for the generic points
will be presented in the following section. These multicritical
points correspond to
deviations from the two extreme limits of vector models and square
matrix models, and only %. These critical points actually yield
%double-scaling solutions since only
one of the large-$N$ parameters, $M$ or $P$, diverges
while the other remains finite. The
recursion equations, presented in section 1, lead to a coupled set of
differential and finite difference equations. The difference equations
relate scaling functions at different values of the finite parameter.
The finite parameter also appears as an extra constant in the differential
scaling equations. Therefore we refer to the results for these critical
points as parametric solutions. We shall show that
the new methods introduced in this paper lead to results that
both include, and extend, known
results\cit{doug4,arley1,arley2,yoneya,div}.

\def\part{\partial}

\def\th{\theta}
For the simplest nontrivial potential, the vanishing of the Jacobian
determinant \(jake) becomes
$$|J| = a_1^2 +
4a_1a_2(\phi_c+\th_c)+4a_2^2(\phi_c^2+\phi_c\th_c+\th_c^2)=0\ .
\eqno(thinair)$$
%As in the previous section,
We normalize $g_c=1=\th_c$ and set $\phi_c=y^2$.
In this case, \(thinair) yields
$$a_1=2{1+y+y^2\over2y+1}\ ;\qquad a_2=-{1\over2y+1}\ .$$
(A second solution, which yields identical results, is produced by
replacing $y\rightarrow -y$.)
Note that none of the individual derivatives appearing in the Jacobian
matrix vanish at this critical point.
The numerator in $a_1$ is a positive definite function of $y$, and it is
straightforward to show $a_1/a_2 \le - 3/2.$
This implies, in particular, that ${\rm sgn}a_1=-{\rm sgn}a_2$ for
critical behaviour. One finds that $q_c=y^3(2+y)/(2y+1)$.
Requiring that $g_c=1\ge q_c\ge 0$, one finds two possible
ranges for $y$: $0\le y\le 1$ and $-2\le y\le -1$. %Note that
Only the former  range satisfies $\th_c\ge \phi_c$,
while only the latter corresponds
to a stable potential. We define $\Delta\th,\Delta\phi,\Delta
g,\Delta q$ as in the previous section, to obtain
$$\eqalign{
\var g &=-{2\over2y+1}(\var\phi-y\var\th)-{1\over2y+1}(\var\th^2+2\var\th
\var\phi)\cr
\var q &={2y\over2y+1}(\var\phi-y\var\th)-{1\over2y+1}(\var\phi^2+2\var\th
\var\phi)\ \ .\cr}\eqno(blow)$$
This singularity has
two notable features: The linear combination $\hat{q}=q+y g$
is distinguished since the expansion begins at quadratic order for this
parameter. Therefore one has ${\part\hat{q}\over\part\th}=0={\part\hat{q}\over
\part\phi}$. Eq. \(blow) also singles out $\hat\phi=\phi-y\th$ as a special
linear combination of the recursion coefficients. This is because the
variations of any linear combination of $g$ and $q$, {\it except for}
$\hat{q}$, about the critical point are linear in $\var\hat{\phi}$. %This
%leaves one to choose the remaining variables to parameterize the critical
%point.
The universal behavior is independent of %unaffected by any specific
the choice of the remaining variables, %and
so it is best made on the basis of simplifying the calculations.
At present,
we wish to focus on exceptional points where the nature of the singularity
changes. In the previous section, we saw that such changes may occur at
the end points in the allowed range of $y$. %We will simply
%claim that
In the
present example that there are two cases of interest: i) $y=0$ and ii) $y=-1$.
This fact will be verified in the %detailed
analysis of the generic points %presented
in sect.~5.

\medskip
\leftline{(i) $y=0:\ \ g_c=1,q_c=0,\th_c=1, \phi_c=0,a_1=2,a_2=-1$}
\smallskip
\par\noindent The fact that $\phi_c=0$ has some important implications.
First from \(bound), the width of the eigenvalue distribution in the saddle
point analysis has shrunk to zero. Thus the analysis given there is
inappropriate for this particular critical point, but it suggests that the
correct saddle point analysis is that given in \cit{arley2}
for vector models. In the saddle point analysis for those models,
$M/\beta\rightarrow 0$ so that the eigenvalues accumulate at a single point
since their electrostatic repulsion is unable to keep them apart.
One can verify that the critical values given above %correctly
correspond to
one of the critical points for the vector models.
This would imply that $M$ should remain finite in the present case.
This conclusion is further supported by the fact that $\al_c=\th_c\phi_c=0$.
Given the usual connection in \(freebird)
of $\al$ to the free energy for a standard
scaling solution, $\al_c=0$ suggests that differences in $M$
do not give rise to %infinitesimal
derivatives suppressed by powers of $\be$ in the scaling limit.

Returning to the expansion about the critical point,
one has $\hat q=q$ and $\hat\phi=\phi$.
It is convenient to choose
$\theta$ and ${P/\beta}\equiv{(N-M)/\beta}=g-q$
as the remaining variables with which to describe the singularity.
%In terms of these variables,
Now \(blow) reduces to
$$\var(g-q)=-2\var\phi-\var\theta^2\ ,\qquad
\var q=-2\,\var\th\var\phi\eqno(blow0)$$
where a term of order $\var\phi^2$ has been dropped from both expressions,
since in the scaling limit
they are higher order. Rewriting \(blow0) as $\var q=\var\th^3+\var\th
\var(g-q)$, one can recognize this singularity as a Whitney pleat\cit{sade}.

\def\be{\beta}
Let us introduce scaling variables as in
\(cordee2,derivative,solve),
$$\eqalign{
{{P}\over\be} &\equiv 1 -\beta^{\nu-1}t,\cr
{{M}\over\be} &\equiv \beta^{\hat\nu-1}s,\cr  }\eqno(where)$$
and
$$\eqalign{
\theta^{(P+l)}_{m} &\equiv 1
-\exp\left[-l\be^{-\nu}\part_t\right]\be^{-\mu} f_{m}(t)\ ,
\cr
\phi^{(P+l)}_{m} &\equiv
-\exp\left[-l\be^{-\nu}\part_t\right]\be^{-\hat\mu}
g_{m}(t)\ .\cr}\eqno(above)$$
Following the previous discussion, we have not introduced $m=M+n$ or
derivatives with respect to $s$ for differences in $m$.
{}From \(blow0), we know the
exponents are related as $\hat\mu=2\mu$, $\nu=1-2\mu$ and
$\hat\nu=1-3\mu,$ so it remains to find $\mu.$  For this, we turn to
\(t1,t2), and find
$$\eqalign{0&=g_m -  g_{m-1} + \part_tf_m,\cr
0 &=g_{m-1}( f_{m-1}-f_m) + {\part_tg_{m-1}},\cr}\eqno(pip2vec)$$
provided that $\mu=\nu,$ which in turn implies $\mu={1\over 3},$
and $\hat\nu=0.$  With the vanishing of this last exponent, one sees that
as expected $M/\beta$ does not scale, but rather \(where) would give
$M=s$. Thus
it is inappropriate to replace discrete differences in the subscripts
in \(above) with
derivatives, since the derivatives would not be suppressed by negative
powers of
$\be,$ and we would have to keep track of arbitrarily high powers of
$\part_s.$

Applying our {\it ans\"atze} to \(dyn1b,dyn2b) to complete the scaling
analysis, we find
$$\eqalign{0 &= m+g_m(f_m+f_{m+1})\cr
0 &=t+g_m+g_{m-1}-f_m^2\ . \cr}\eqno(pip3)$$
Combining these results with \(pip2vec), we derive a single
equation in terms of $f_m$ only, which takes the surprising form
$$\part_t^2f_m-2f_m^3+2tf_m=2m-1\ .\eqno(polymer)$$
This is the Painlev\'e II equation, with a constant, and it
also
arises in the scaling solution of unitary matrix models with
boundaries\cit{joe1}.  Its solutions have been
studied in detail as well\cit{joe2joe3}. Note that in the present case,
the constant appearing on the right hand side of \(polymer) is an odd
positive integer, while the constant in \cit{joe1,joe2joe3} is an arbitrary
integer corresponding to the number of quark flavors coupled to the
models.

The above results are valid for any $m\ge1$. %Note though
{}From \(cur2), one has $\rp l0=0,$ so  $g_0=0$. Thus at $m=1$,
\(pip2vec) and \(pip3) give: $0=g_1+\part_tf_1$ and $0=t+g_1-f_1^2$.
Therefore $f_1$ satisfies
$$ t=\part_tf_1+f_1^2,\eqno(short)$$
the $k=2$ scaling equation in the hierarchy arising from vector
models\cit{arley2}. This result is consistent with
\(polymer),  which can be written as
$$(\part_t-2f_m)(\part_tf_m+f_m^2-t)=2(m-1)\ \ .$$
Hence at $m=1$, \(polymer) contains the second vector equation \(short).

Since $M$ does not scale, the relation of $f_m$ to the free energy is slightly
more complicated than in other matrix
models. One has
$$\exp(-\part_PF_M)\approx{\cz_{\ssc M,P+1}\over\cz_{\ssc M,P}}=
\prod_{m=1}^M{\h {P+1}{m-1}\over\h {P}{m-1}}=\prod_{m=1}^M\rt Pm$$
which leads to
$$\part_t F_M=-\sum_{m=1}^Mf_m\ \ .\eqno(sumfree)$$
Given the asymptotic expansion arising from \(polymer)
$$f_m=t^{1\over2}-{2m-1\over4}t^{-1}-{12m(m-1)+5\over32}t^{-{5\over2}}-
\ldots\ ,$$
one finds
$$\part_t F_M=-M\,t^{1\over2}+{M^2\over4}t^{-1}
+{4M^3+M\over32}t^{-{5\over2}}+\ldots$$
These results agree with those  for
the vector models using the analysis given in \cit{arley2}.  One finds,
therefore, an {\it a priori} surprising connection between solutions of
\(polymer), and the Wronskian of the flows of Burger's
equation\cit{arley2}.

Therefore we have confirmed that the scaling behavior of this critical point
corresponds to a particular
scaling solution of the vector models, which describe
dynamically branched polymers. This case generalizes to the complete
hierarchy of vector critical points, which are characterized by
$\phi_c=0=q_c$. Without giving the full analysis, we give the solution for
the next critical point in this series.
When $g_c=1$, $q_c=0$, $\th_c=1$, $\phi_c=0$,
$a_1=-a_2=3, a_3=1,$ we find $\mu=\nu={1\over 4},\hat\mu={1\over
2},\hat\nu=0,$ and
$$\eqalign{0&=m+g_m(f_m^2+f_mf_{m+1}+f_{m+1}^2-g_{m-1}-g_m-g_{m+1})\cr
0&=t-f_m^3+f_{m-1}g_{m-1}+g_mf_{m+1}+2f_m(g_m+g_{m-1})\ \ .\cr}$$
Eq.~\(pip2vec) remains unchanged, and combined with the above
results, we find
$$\eqalign{
0=t^2f_m&+6(1-2m)f_m^3+(4t-5f_m^3)f_m^4-4f_m
\part_tf_m\cr
&+(4t+5f_m^3)(\part_tf_m)^2 -(4t-10f_m^3)f_m\part_t^2f_m
-4(\part_tf_m)^2\part_t^2f_m\cr
&+3f_m(\part_t^2f_m)^2+4f_m\part_tf_m
\part_t^3f_m-2f_m^2\part_t^4f_m\ .\cr}$$
One also finds that at $m=1$
$$t=f_1^3+3f_1\part_tf_1+\part_t^2f_1$$
which is the $k=3$ vector scaling
equation. There is also a topological
(\ie nonsingular) scaling solution for $g_c=1$, $q_c=0$, $\th_c=1$,
$\phi_c=0$,
$a_1=1$, with $\mu=\nu={1\over 2},\hat\mu={1},\hat\nu=0,$ and
$g_m=-m,\ f_m=t$. In all cases, \(sumfree) determines the free energy given
the scaling solutions for $f_m$.

\medskip
\leftline{(ii) $y=-1:\ \ g_c=1,q_c=1,\th_c=1,\phi_c=1,a_1=-2,a_2=1$}
\smallskip
\par\noindent We find very similar scaling behavior at this critical point
despite the fact that it describes very different physics.
This critical point involves square matrices since
$(P/\beta)_c=g_c-q_c=0,$ though they may deviate from being precisely square
in the scaling limit. Since $\th_c-\phi_c=0$, \(bound) indicates that
the critical behavior involves the approach of the inner limit of the
eigenvalue distribution to the origin. For $P=0$ exactly, a series of
scaling solutions for such critical points have been studied in
Ref.~\cite{timstuff}.

Considering the expansion about the critical point, one has $\hat{q}=
q_c-g_c=-P/\beta$ and $\hat{\phi}=\phi+\theta$. A convenient choice of
variables to complete the description of the singularity are $q$ and
$\th-\phi$. Defining $\var_{(+)}\equiv\var(\theta+\phi)$ and $\var_{(-)}
\equiv\var(\th-\phi)$, \(blow) becomes
$$
\var(g-q)=\var_{(+)}\var_{(-)}\ ,\qquad\var q=2\var_{(+)}-{1\over4}
\var_{(-)}^2 \eqno(blow1)$$
where we have dropped terms that are higher order in the scaling limit.
Rewriting \(blow1) as $\var(g-q)={1\over4}\var_{(-)}^3+{1\over2}
\var_{(-)}\,\var q$,
one recognizes that this critical point is again characterized by
a Whitney pleat\cit{sade}.
\def\del{\delta}
\def\k#1#2{k^{(#1)}_{#2}}

We find a scaling solution similar to that for case (i).
Defining $\del\equiv\beta^{-{1\over 3}}$, we set
$$\eqalign{
{P\over\beta}&=\del^3\,t\cr
{M\over\beta}&=1-\del^2\,s\ \ ,\cr}\eqno(where1)$$
and
$$\eqalign{
\rt l{M+n}+\rp l{M+n}&=2-2\exp[-n\del\part_s]\,(\del^2\h l2(s)+\del^3\h l3(s)
+\ldots)\cr
\rt l{M+n}-\rp l{M+n}&=-2\exp[-n\del\part_s]\,(\del^1\k l1(s)+\del^2\k l2(s)
+\ldots)\ \ .\cr}\eqno(above1)$$
Hence in this case, $P$ does not scale, so no derivatives
with respect to $t$ are introduced. Eq.'s \(dyn1b,dyn2b) produce
scaling solutions for fixed $l$. One finds
$$\eqalign{
\h l2&={1\over4}(s-\k l1{}^2+\part_s\k l1),\cr
\h l3&=-{1\over8}(l+4\k l1\,\k l2-2\part_s\k l2)\ \ .\cr}$$
The scaling function $\k l1$ satisfies
$$\part_s^2\k l1-2\k l1{}^3+2s \k l1 = 2l+1,\eqno(poly2)$$
which is the Painlev\'e II equation with a constant, once again.
The scaling parameters associated with $M$ and $P$ have traded r\^oles
in \(poly2), as compared to \(polymer). Since $M/\beta$ scales in the
present case, \(freebird) allows one to determine the free energy in terms of
$\k l1$ as
$$\part_s^2F^{(P)}=-{1\over2}(s+\k P1{}^2+\part_s\k P1)\ \ .\eqno(freeone)$$
For this critical point, the potential independent results derived
from \(t1,t2) play an
auxiliary r\^ole, relating the scaling function at different values of $l$
$$\k l1{}^2-\part_s\k l1=\k {l+1}1{}^2+\part_s\k {l+1}1\ \ .$$
Thus the free energies at different values of $P$ will also be related for
these models.

The above results remain unchanged at $l$ or $P=0$. This corresponds to the
case of precisely square matrices, which were studied in Ref.~\cite{timstuff}.
As noted in case (i), one can write \(poly2) with $l=0$ as
$$(\part_s-2\k 01)(\part_s \k 01+\k 01{}^2-s)=0\ \ .$$
Thus the conventional solution of \(poly2) with the asymptotic expansion
$\k 01=\pm s^{1/2}+\ldots$, leads to a trivial free energy $\part_s^2F^{(0)}
=-s$ with no higher order terms. Fortunately, the Painlev\'e II equation
with a nonvanishing constant admits a third solution, which for $l=0$ is
$$\k 01={1\over2}s^{-1}-{3\over8}s^{-4}+{111\over32}s^{-7}-\ldots\ ,$$
yielding
$$\part_s^2F^{(0)}=-{1\over2}s+{1\over8}s^{-2}-{9\over16}s^{-5}
+{1323\over128}s^{-8}-\ldots\ \ .$$
We have checked that up to
rescaling variables, the universal (\ie nonanalytic)
part of the free energy is identical to that found for the
first critical model in Ref.~\cite{timstuff}. Note that the latter analysis
describes the free energy in terms of a scaling function, which satisfies
Painlev\'e II without a constant.

This case generalizes to a hierarchy of critical points characterized
by $\th_c=\phi_c$ and $g_c=q_c$. At $P=0$, our analysis will reproduce
the results of Ref.~\cite{timstuff}. There a new series of scaling solutions
describing random surfaces, was
constructed. Their asymptotic expansions remain governed by the
conventional differential
equations arising in the analysis of Hermitian matrix models\cit{big3}.
For $P\not=0$,
the matrices in the present models, deviate slightly from
being square. The scaling solutions of these matrix models are the limit
II solutions discussed in \cit{arley1}. The relation between the present
analysis and that given in the previous references for both these
random surface and the previous vector critical points is under
investigation\cit{lafrance}.

\section{5. Scaling Solutions with Partial Differential Equations}

We now return, for generic values of $y$, to \(blow) which gave the
expansion about the critical points associated with the simplest
nontrivial potential. Along with $\hat \phi=(\phi
-y\th)$ and $\hat q=(q+yg)$, we choose $\th$ and $g$ to give a
representative parameterization of the singularity, \(blow) becomes
$$\eqalign{
\var g&=-{2\over2y+1}\var\hat{\phi}-\var\th^2-{2\over2y+1}\var\hat{\phi}
\var\th\cr
\var\hat{q}&= -2\var\hat{\phi}\var\th-{1\over2y+1}(\var\hat{\phi}^2
+3y(y+1)\var\th^2)\ \ .\cr}\eqno(neqone)$$
Now as long as $y(y+1)\not=0$, one can define $\hat{g}=g-
{(2y+1)\hat q/3y(y+1)}$ for which the relevant expansion is simply
$$\var\hat{g}=-{2\over2y+1}\var\hat{\phi},\eqno(linear)$$
since the remaining quadratic terms contain $\var\hat{\phi}$ and hence
are higher order in the scaling limit. Note that \(linear) is the complete
expansion. Cubic or higher order terms are simply not possible since
the potential $V(y)$ is only quadratic. Choosing
$\hat{\th}=\th+{(2y+1)\hat{\phi}/3y(y+1)}$, the $\hat{q}$ expansion
becomes
$$\var\hat{q}=-{1\over2y+1}\left(3y(y+1)\ \var\hat{\th}^2-
{y^2+y+1\over3y(y+1)}\ \var\hat{\phi}^2\right)\ \ .\eqno(quadratic)$$
Eq.'s \(linear,quadratic) characterize the generic singularities. Clearly
this analysis is inappropriate for $y=0$ and $-1$, which verifies the claim
in the previous section that these are higher order multicritical
points where the form of
the singularity changes.

The quadratic form of the singularity in \(quadratic), might lead one to
conjecture that $\var\hat{q}\approx-\del^4\,t$ where $\del=\beta^{-{1
\over 5}}$.
The exponents are chosen here
in analogy to those which characterize the Painlev\'e
I equation, which arises from a quadratic singularity in the standard
double-scaling analysis. Given this choice, \(linear) would lead
to $\var\hat{g}\approx-\del^2\,s$. Discrete differences in the large-$N$
parameters, $\be\hat{g}$ and $\be\hat{q}$, would
then give rise to %infinitesimal
derivatives, $\be^{-\nu}\part_t=\del\part_t$ and $\be^{-\hat{\nu}}\part_s
=\del^3\part_s$, respectively. We will
find that these choices do indeed lead
to a triple-scaling solution below. An essential feature of this
solution is the ratio $\hnu/\nu=3$.
We shall show that with this choice, a partial differential equation very
similar to the full time-dependent KdV equation
is obtained from \(t1,t2), without any assumptions restricting $V(y).$
Suitable tunings of  $V(y)$ lead to a hierarchy of additional differential
equations.  Each of these new critical points is governed by
a system of two partial differential equations, one of which is the
KdV-like equation.

\subsection{5.1 Potential Independent Relations}

In accord with \(cordee), we begin with a general scaling {\it ansatz} for
two linearly independent combinations of $M$ and $P$
$$\eqalign{
{A\,M+C\,P\over\beta}&=\left[{A\,M+C\,P\over\beta}\right]_c-\beta^{\nu-1}t\cr
{B\,M+D\,P\over\beta}&=\left[{B\,M+D\,P\over\beta}\right]_c-\beta^{\hnu-1}s
\ .\cr}\eqno(map)$$
In the scaling limit, discrete differences become derivatives
as follows:
$$\part_M=-\be^{-\nu} A\part_t-\be^{-\hnu}B\part_s\qquad{\rm and}
\qquad\part_P=-\be^{-\nu} C\part_t-\be^{-\hnu}D\part_s\ \ .
\eqno(derder)$$
Now set $\hnu/\nu=3$ as above, and define $\del\equiv\be^{-\nu}$.
We also choose $\th_c=1$ and $\phi_c=y^2$ as in previous sections, and
then introduce
$$\eqalign{
\rt{P+l}{M+n}&=1-\exp\left[-\del\left(An+Cl\right)\part_t
-\del^3\left(Bn+Dl\right)\part_s\right]\ \sum_{q=2}\del^qh_q(t,s),\cr
\rp{P+l}{M+n}&=y^2-\exp\left[-\del\left(An+Cl\right)\part_t
-\del^3\left(Bn+Dl\right)\part_s\right]\ \sum_{q=2}\del^qk_q(t,s)\ \ .\cr}
\eqno(orwhat)$$
%In beginning the two series
%in \(orwhat) at $q=2$, we use the relations
%of the recursion coefficients to derivatives of the free energy derived
%from \(freebird).

Substituting \(orwhat)\ into \(t1,t2), we find
$$y= \pm{C\over A-C}\ .\eqno(result1)$$
One may choose the plus sign here without any loss of generality.
Then
$$\eqalign{
k_2(t,s) &= y\left(h_2(t,s) + g_2(s)\right),\cr
k_3(t,s) &= y\left(h_3(t,s) - {A\over2}\part_th_2(t,s)
+ g_3(s)\right) ,\cr}\eqno(result2)$$
and finally, one solves for $k_4$ at $O(\del^5)$, but consistency between
\(t1) and \(t2) requires
$$ \part_sh_2={C^2\over4T}{y-1\over y^2}\ \part_t\!\left[h_2{}^2-{C^2\over6}
{y+1\over y}\part_t^2h_2\right]-{C^2\over2Ty^2}g_2\part_th_2
-{1\over2}\part_sg_2 \eqno(kdv)$$
where $T=A\,D-B\,C$. %The latter is guaranteed to be nonvanishing since
$T$ is the determinant of the linear transformation relating $\var(M/\be)$
and $\var(P/\be)$ to the scaling variables, $t$ and $s$, in \(map), so
we require that $T\not= 0.$
Neglecting the last two terms involving $g_2$, \(kdv)
is the KdV equation
with $s$ and $t$ playing the r\^oles of the time and space coordinates,
respectively. We stress that this result was obtained only assuming
a particular ratio of the exponents in \(map) and \(orwhat)
(\ie $\hnu/\nu=3$ and $q\ge2$), %and
without restricting the matrix potential in any way.

\subsection{5.2 Potential Dependent Results for Quadratic Potential}

We return to discussing the generic critical point associated with
the quadratic potential. For a scaling solution of
\(dyn1b,dyn2b), one finds $\nu={1\over 5}$,
$$g_2(s)=-{C\over2T}{2y+1\over y^2} s\qquad{\rm and}\qquad g_3(s)=0\ \ .
\eqno(simple)$$
Finally $h_2$ must satisfy the following differential equation
$$0=t-{C\,y\over2y+1}g_2{}^2-2C g_2h_2-3C{y+1\over2y+1}\left[
h_2{}^2-{C^2\over6}{y+1\over y}\part_t^2h_2\right]\ \ .
\eqno(paine)$$
This equation has a form similar to the Painlev\'e I equation, again up
to terms involving $g_2$, which introduce $s$ dependence.

Up until this point, we have not made use of the freedom available to
rescale the variables: $s\rightarrow\varpi s$ and $t\rightarrow\varphi t$;
and the functions: $h_2\raro\gamma h_2$ and $g_2\raro\eta g_2$;
to simplify our results. Note that these rescalings do {\it not} provide
enough freedom to completely eliminate $y$ from the coefficients
in \(kdv,simple,paine). A suitable rescaling brings these equations to
the following form: $g_2=s$;
%$${\be\over\ga^2}=3C{y+1\over2y+1};\be^2\ga={C^2\over27}{y+1\over y};
%{\al^2\over\be}={12T^2y^4(y+1)\over C^3(2y+1)(y^2+y+1)};{\delta\over
%\al}=-{C\over2T}{2y+1\over y^2}$$
$$\eqalign{
0&=t-{3y(y+1)\over y^2+y+1}g_2{}^2+2{2y+1\over\sqrt{y^2+y+1}}g_2h_2
-h_2{}^2+{9\over2}\part_t^2h_2\ ;\cr
\part_sh_2&={y-1\over2\sqrt{y^2+y+1}}\part_t
\left[h_2{}^2-{9\over2}\part_t^2h_2
\right]+{3(y+1)\over y^2+y+1}g_2\part_th_2+{3(y+1)\over2\sqrt{y^2+y+1}}
\part_sg_2\ .\cr}$$
Although $C$ and $T$ have been eliminated from the coefficients,
the last two equations still appear rather complicated.
The true simplicity of our results is made manifest by
applying $g_2=s$, and shifting $h_2=\hat{h}+{(2y+1)s/\sqrt{y^2+y+1}}$,
which then yields
$$\eqalign{
0&=(t+s^2)-\hat{h}^2+{9\over2}\part_t^2\hat{h}\cr
0&=\part_s\hat{h}-2s\part_t\hat{h}\cr}\eqno(really)$$
where $\part^3_t\hat{h}$ has also been eliminated from the KdV-like equation,
by using the first equation above.
Thus \(really) shows that all values of $y$ (which indirectly determines the
proportions of the matrices) give the same scaling behavior.
Note though that the shift between $h_2$ and $\hat{h}$
introduces a separate $s$ dependence with a coefficient involving $y$.
%This suggests that this $s$ dependence is in fact nonuniversal, as
%we will discuss further next.

Given the scaling equations \(really),
there are some complications in determining
the free energy. Eq.~\(derder) shows that both $\part_M$ and $\part_P$
are both proportional to $\part_t$, to leading order.
{}From \(freebird), the expression
for $\ra PM$ yields
$$\part_t^2F\approx-{1\over\de^2}{2y^2\log y\over C^2(y+1)^2}
+{y\over C^2(y+1)}h_2 + {y\over C^2(y+1)^2} g_2\ \ , \eqno(thisone)$$
while that for $\rt PM$ yields
$$\part_t^2F\approx{y\over C^2(y+1)}h_2\ \ .\eqno(thistwo)$$
The disagreement between these expressions arises because the
different approximations introduce different nonuniversal contributions.
The divergent constant explicitly included in \(thisone) is a familiar
term, which is regularly dropped (as we did implicitly in section 3).
The finite difference between \(thisone) and \(thistwo) proportional
to $g_2(s)$ may be less familiar, but is simply another nonuniversal
contribution. Consistent expressions for $\part_M\part_P^2 F$, $\part_M^2
\part_P F$ and $\part_M^3 F$ can be produced by taking appropriate
ratios of the recursion coefficients $\theta$ and $\al$, with the result:
$$\part_t^3F\approx{y\over C^2(y+1)}\part_th_2.$$
Therefore up to rescalings
which can be absorbed in $F$, we define the universal part of the free
energy as
$$F(t,s)=\int^t\!dt_1\int^{t_1}\!dt_2\ \hat{h}(t_2,s)\eqno(universe)$$
where any possible analytic dependence on $t$ is dropped, as well as any
separate $s$ dependence.\footnote{$^\dagger$}{One might think that %
constructing $\part_t\part_s F$ or $\part_s^2 F$ would reveal the `correct' %
$s$ dependence. %
We %pursued this question to a certain extent, but
were unable to find %
consistent definitions in general. For instance, one might expect that %
$\part_t\part_s F$ will appear to leading order in $\part_M(C\part_M-A %
\part_P)F$, but either $\log\left[(\ra PM)^C/(\rt PM)^A\right]$ or %
$\log\left[(\ra PM)^C/(\rt P{M+1})^A\right]$ appear to be equally good %
approximations for this quantity. %
Unfortunately the two expressions differ at the subleading %
order $\del^3$, while $\part_t\part_s F$ appears at order $\del^4$.}

Now consider solving the scaling equations \(really). The second equation is
simply solved by $\hat{h}=\hat{h}(t+s^2)$. Given this form, the first
equation can be solved by a perturbative expansion identical to that
constructed for the Painlev\'e I equation, where the expansion parameter
is now $(t+s^2)^{-5/2}$. Thus one finds the free energy is given by
$$\eqalign{
F=&{4\over15}(t+s^2)^{5\over2}+{9\over16}{\rm ln}(t+s^2)-
{567\over640}(t+s^2)^{-{5\over2}}+\ldots\cr
=&{4\over15}t^{5\over2}\left(1+{5\over2}{s^2\over t}+{15\over8}
{s^4\over t^2}+\ldots\right)\cr
&+{9\over16}{\rm ln}(t) +\left({9\over16}
{s^2\over t}-{9\over32}{s^4\over t^2}
+\ldots\right)\cr
&-{567\over640} t^{-{5\over2}}
\left(1-{5\over2}{s^2\over t}+{35\over8}{s^4\over t^2}+\ldots\right)+
\ldots\cr}
\eqno(expando)$$
These results can be interpreted as follows: The free energy has
the conventional
topological expansion in terms of $t^{-{5\over 2}}$,
but the contribution of each
topology is perturbed by terms $({s^2/t})^m=(s/t^{{1\over2}})^{2m}$.
In the present case,
these perturbations can be resummed to give an overall
renormalization of the
topological expansion parameter of the form $(t+s^2)^{-{5\over 2}}$.

It is obvious from \(expando)\
that there is a dual asymptotic solution, valid when $s$
is large and $t$ is small.
This dual expansion merely has $t^{5\over 2}$
replaced with $s^5,$ with $s^2/t \rightarrow t/s^2.$
In the next section, we shall find that the dual expansion has a rather
distinct character---neverthelsss, the important point is the
existence of a dual expansion for these critical points.

\subsection{5.3 Higher Order Singularities}

This section discusses some results for the higher order multicritical
models, for which the KdV-like equation governs the scaling behavior.
In section 3, we described how a matrix potential with
$(2n-1)$ nonvanishing coefficients could be tuned to to yield a
split singularity
of the form $y\var g\pm\var q\propto(y\var\th\pm\var\phi)^n$.
One extra nonvanishing coefficient (for a total of $2n$) will allow
for extra tuning so that $y\var g+\var q$ can be made to vanish to order
$n$ in the variations, $\var\th$ and
$\var\phi$.\footnote{$^\ddagger$}{Clearly we could %
have also singled out $y\var g-\var q$, but this yields the same results.}
This construction leads to the singularities where the scaling solution
of section 5.1 is relevant.

Explicitly for the case $n=2$, the minimal potential has $a_1$, $a_2$, $a_3$
and $a_4$ nonvanishing. We choose $g_c=1=\th_c$ and $\phi_c=y^2$. Then
setting
$$\eqalign{
a_1&=-2(2y^6+3y^5+3y+2)a_4\cr
a_2&=6(y^4+y^3+y^2+y+1)a_4\cr
a_3&=-2(2y^2+y+2)a_4\cr
a_4&={1\over6y^3+2y^2-2y-1}\cr}\eqno(neqtwo)$$
yields a singularity of the form
$$\eqalign{
\var g&\propto \var\hat{\phi}^2\cr
\var\hat{q}&\propto b_3\,\var\hat{\phi}^3 + b_2\,\var\hat{\phi}^2\var\th
+b_1\,\var\hat{\phi}\var\th^2+\var\th^3\cr}$$
where the notation is the same as used in \(neqone), and $b_i$ are
constants depending on $y$. Note that in this case unlike \(quadratic),
one cannot choose a
linear combination $\hat{\th}=\th+x\hat{\phi}$ to eliminate all of the
terms mixing $\var\hat{\phi}$ and $\var\hat{\th}$ in the singularity,
although the term linear in $\var\hat{\phi}$ could be removed.
For \(neqtwo), one finds that $q_c=y^5(6+2y-2y^2-y^3)/(6y^3+2y^2-2y-1)$.
Requiring $g_c=1>q_c>0$, leaves two possible ranges for $y$: $-1<y<0$ or
$1<y<\xi$ with $\xi=-2/3+10^{1/3}/3+100^{1/3}/3\approx1.59867$.

Now one may investigate a scaling solution with the {\it ansatz} given
in section 5.1. The results presented there \(result1,result2,kdv),
are of course unchanged. A scaling solution of \(dyn1b,dyn2b)
requires
$$\eqalign{
\nu&={1\over7}\ ;\qquad g_2=s^{{1\over2}}\ ;\qquad g_3=0\ ;\cr
0&=t-h_2{}^3+h_2\part_t^2h_2+{1\over2}(\part_th_2)^2-{1\over10}\part_t^4h_2\cr
&\quad+{1\over2}g_2\part_t^2h_2-{3\over2}g_2h_2{}^2-{3\over5}
{y^2+3y+1\over(y+1)^2}g_2{}^2h_2
-{1\over10}{y(y+4)\over(y+1)^2}g_2{}^3\cr}$$
where rescalings have been made to simplify these results. A further
simplification comes from shifting $h_2=\hat{h}-\half g_2,$ applying
$g_2=s^{{1\over2}}$, and making the following definitions:
$$x={y-1\over y+1}\ ,\qquad \tau=t-{x\over20}s^{3\over2}\ ,\qquad
\sigma={3x^2\over20}s\ ,$$
which yields
$$0=\tau-\hat{h}^3+\hat{h}{\part}_\tau^2\hat{h}+{1\over2}(\part_\tau
\hat{h})^2-{1\over10}{\part}_\tau^4\hat{h}+\sigma\,\hat{h}
\ \ .\eqno(kdv21)$$
Neglecting the last term involving $\sigma$, this is the
second equation in the hierarchy that arises for scaling
solutions of hermitian matrix models\cit{big3}.
With the corresponding rescalings and the shift of $h_2$,
\(kdv) becomes the KdV equation
$$\part_\sigma\hat{h}={1\over2}\,\part_\tau\!\left[
\hat{h}^2-{1\over3}{\part}_\tau^2\hat{h}\right]\ \ .\eqno(kdv2)$$
Note that all of the $y$ dependence has been absorbed in the definitions of
$\tau$ and $\sigma$, leaving the scaling behavior independent of the
proportions of the matrices, once again.

One can construct a perturbative solution  of these scaling
equations for large $\tau :$
$$\eqalign{
F=&{9\over 28}\tau^{7\over3}
\left[1+ {14\over 15}{\sigma\over\tau^{2\over3}}+
{14\over 81}{\sigma^3\over\tau^{2}} + {7\over 243}
{\sigma^4\over\tau^{8\over3}}
+\dots\right]\cr
&+{1\over 18}{\rm ln}\tau + \left[{1\over 36}{\sigma\over\tau^{2\over3}}
+{1\over 216}{\sigma^2\over\tau^{4\over3}}
-{1\over 243}{\sigma^3\over\tau^{2}}+
{11\over 11664}{\sigma^4\over\tau^{8\over3}}
+ \dots\right]\cr
&- {1\over 120}{\tau}^{-{7\over 3}}\left[1+
{25\over 81}{\sigma\over\tau^{2\over3}}
-{170\over 243}{\sigma^2\over\tau^{4\over3}}
+{505\over 729}{\sigma^3\over\tau^{2}}+
\ldots\right]\quad +\ldots\cr}\eqno(expkdv2)$$
These results are more complicated than the previous case, \(expando):
a conventional topological expansion can be constructed
in terms of $t^{-{7\over3}}$, dressed at each order by
perturbations $(s^{{1\over 2}}/ t^{{1\over3}})^m$. The odd powers
can be absorbed in an overall renormalization of the expansion
parameter, $t\rightarrow\tau,$ as seen explicitly in \(expkdv2).
Perturbations with
even powers remain in \(expkdv2), and can not be incorporated into
such a renormalization.
In part, the latter complication arises because
eq.~\(kdv2) cannot be reduced to
a linear equation as in \(really). Perhaps then one should think of
\(really) as the `topological point' for the physics described by
these partial differential equations.

\def\ter#1#2#3{{\tau^#1\over{\sigma^{{#2\over #3}}}}}
The dual expansion at this critical point (for ${\sigma^3\gg\tau^2}$) is
$$\eqalign{\hat h = & \sigma^{{1\over 2}}
\bigg[1 + {1\over 2}
{\tau\over{\sigma^{{3\over 2}}}}
- {3\over 8}
{\tau^2\over \sigma^3}
+{1\over 2}\ter392 - {105\over 128}
{\tau^4\over \sigma^6}
+ \dots\bigg]\cr
&-{5\over16}\sigma^{-3}\bigg[1-{51\over 10}
{\tau\over{\sigma^{{3\over 2}}}}
+{393\over 20}
{\tau^2\over \sigma^3}
-{5403\over 80}\ter392+{3489\over 16}
{\tau^4\over \sigma^6}
+ \dots\bigg]\cr
&-{2451\over 512}\sigma^{-{13\over2}}\bigg[1-{9406\over 817}
{\tau\over{\sigma^{{3\over 2}}}}
+{531501\over 6536}
{\tau^2\over \sigma^3}
-{741305\over 1634}\ter392
+{229444887\over 104576}
{\tau^4\over \sigma^6}
+\dots\bigg]\quad +\dots\cr}\eqno(expkdv3)$$
Again, note that it is not possible to absorb the perturbations, now
in powers of $\tau \sigma^{-{3\over 2}},$ into a renormalization of
$\sigma.$  There is an analogous series with $\hat h \sim -
\sigma^{1\over 2} +\dots.$
We have given here the expansion of the specific heat, not the free
energy, because our methods (as mentioned in a preceding footnote) do
not enable us to unambiguously determine the derivatives of the
free energy with respect to $s.$  Thus, we are not able to
determine from the triple-scaling equations, contributions
which are non-analytic in $s$ but have a $\tau$ dependence of the
form $a+b\tau.$  It would be a tedious but straightforward exercise to
obtain these terms using the techniques of \cit{biz}.
Note that the dual expansions \(expkdv2,expkdv3)\
are in terms of the renormalized parameters $\tau,\sigma.$
Similar expansions in terms of the bare parameters $t,s$ can
be constructed, as in \(expando).
%The above expansion, \(expkdv3),
%in powers of $\tau$ could be replaced by an expansion in powers of $t,$
%with different coefficients, but an entirely similar form.

There are instanton solutions associated with each of these
weak-coupling expansions.  Linearizing \(kdv21,kdv2) about \(expkdv2),
we find solutions of the form
$$\varepsilon(\sigma,\tau) = \tau^{-{1\over 4}}\left(1+a_{\ssc 1}
{\sigma\over\tau^{2\over3}}
+\dots \right)\exp\left[-{6\over
7}\tau^{{7\over 6}}\left\{r_{\ssc 0} + r_{\ssc 1}
{\sigma\over\tau^{2\over3}} +\dots\right\}\right],$$
where $r_{\ssc 0}:30 -10r_{\ssc 0}^2+r_{\ssc 0}^4 =0,$
$r_{\ssc 1}= -7r_{\ssc
0}(r_{\ssc 0}^2 -6)/36,$ and $a_{\ssc 1} = (r_{\ssc 0}^2 + 2)/24.$
At $\sigma=0,$ these solutions reduce to those found for the
$k=3$ Hermitian matrix model\cit{big3}.

Similarly, linearizing about
\(expkdv3), we find solutions of the form
$$\epsilon(\sigma,\tau) = \sigma^{-{3\over 8}}\left(1+a_{\ssc 1}
{\tau\over\sigma^{3\over2}}
+\dots \right)\exp\left[-{4\over
7}\sigma^{{7\over 4}}\left\{r_{\ssc 0} + r_{\ssc 1}
{\tau\over\sigma^{3\over2}}
+\dots\right\}\right],$$
where now $r_{\ssc 1}:12005 -1960 r_{\ssc 1}^2 + 64 r_{\ssc 1}^4,$
$r_{\ssc 0} = 4r_{\ssc 1}(147 - 8r_{\ssc 1}^2)/1029,$ and
$a_{\ssc 1} = r_{\ssc 1}^2/49 -1/4.$  Note that in both instances,
the prefactor and the exponential are `dressed' by series in
the same perturbing parameters as \(expkdv2,expkdv3).
We have exhibited instantons for this critical point to
display this  `dressing'. The
instantons for the
other critical points found in this paper do not
differ from the double-scaling analysis instantons.

Analogous behaviour is found for higher multicritical points in this
series.  For
general $n$, the singularity may be put in the form
$$\eqalign{
\var g&\propto \var\hat{\phi}^n\cr
\var\hat{q}&\propto \var\hat{\th}^{n+1}+ b_2\var\hat{\phi}^2
\var\hat{\th}^{n-1}+\ldots
\cr}$$
where `$\ldots$' indicates terms higher order in $\var\hat{\phi}$.
In the scaling solution, $g_2=s^{{1\over n}}$ while $\hat{h}$ satisfies
a nonlinear equation similar to the $k=n+1$ equation in the usual
hermitian matrix model hierarchy. This equation will yield the
conventional topological expansion in $t^{-2-{1\over n+1}}$, but
perturbed by terms of the form $\left(s^{1\over n}/t^{1\over n+1}
\right)^m$, as is evident by examining the planar limit. Finally,
these perturbations cannot be expressed as a simple renormalization of
the topological expansion parameter.

\section{6. Conclusions}

It is usually believed that the character of the large-$N$ limit is
different for theories with dynamical variables transforming in
vector-like represenations, or in the adjoint representation.  The
triple-scaling analysis given in the present work shows clearly
that, at least for $d=0$ models, this perceived difference is
a reflection of the limitations of a case-by-case analysis.
We have shown explicitly that
critical phenomena ranging from polymers to random surfaces
are described by the same equations.   Random surfaces are geometrically
associated with triangulations defined by dual graphs of Feynman
diagrams\cit{thooft}, while
polymers are directly associated with the Feynman graphs\cit{arley2}.
One intriguing  result is the appearance of the
second Painlev\'e equation in a model of polymers.
For $k$ even, one might expect that the higher equations should be related
to the higher mKdV flows found in unitary matrix models---this will be
discussed elsewhere\cit{lafrance}.

We have endeavoured, in the present work, to provide a detailed
account of the various subtleties that arise in the triple-scaling
analysis.  The
methods given in this paper seem inappropriate
for understanding the relation, if any, of the multicritical points
of triple-scaling models to hierarchies of integrable equations
{\it e.g.}, {\it \`a la} Drin'feld-Sokolov.
Why should one expect such a relation?  Firstly,
the systems of equations obtained  have a quasi-one-dimensional character.
Secondly, a detailed examination of the examples solved in this paper
will lead to an appreciation of the miraculous cancellations that take
place in order for scaling behaviour in {\it both} variables to
occur---thus, integrability
of some sort must underlie the triple-scaling physics.
The correct mathematical framework should, presumably,
encompass the coupled differential equation-difference equation systems
we found in sect.~4.

We found remarkable dual weak-coupling expansions for the
examples in sect.~5.  The transition from
the domain of convergence of the sub-series associated with one
expansion, to the domain of convergence of the  `dual' expansion,
is of great physical interest---to understand
this, note that small $s$ in the large $t$ expansion, \(expkdv2),
corresponds to {\it strong} coupling in the coupling constant
associated with $s,$ and vice versa in \(expkdv3).  Of course, it
is necessary to solve the problem
of the surface interpretation of the continuum theories before such
a dual relationship can be exploited effectively.

We now suggest an interpretation of the physics of
the new `coupling constant' associated with the $s$ variable.
The Liouville mode is the dynamical degree of freedom
of induced two-dimensional gravity in background gauge.
It is naturally associated with the string coupling constant,
because of the linear dilaton-like\cit{rdil} coupling of the Liouville
field to the background curvature.
In light of eq.~\(expkdv3), it is  natural to conjecture
the existence of a {\it new} dynamical field associated with the
new expansion parameter.  What might the physics of this new
field be?  Based on the surface interpretations that are
suggested by finite $M,N$ computations\cit{arley1,arley2},
the new dynamics
that can be investigated in the triple-scaling analysis would appear to
involve microscopic topology fluctuations.

How might this happen?  The saddle
point analysis\cit{bipz} computes the physics
of random surfaces of spherical topology.  Double-scaling\cit{big3}
solves the physics of random surfaces to all orders in the
topological expansion.  The finer analysis provided by triple-scaling
may then reveal more intricate sub-structure in the physics of
random surfaces.

So much for that\cit{gs}.

\section{Appendix A. Another Higher Order Singularity}

At the end of section 3, we mentioned that when the
split singularity degenerates at $y^2=1$, the scaling solution is similar
to that found in section 5. We will briefly outline these results.
With $y^2=1$, one has $g_c=1=q_c=\th_c=\phi_c$ along with
$a_1=-3$, $a_2=3$, $a_3=-{1\over2}$.
Defining $\var_{(+)}\equiv\var(\theta+\phi)$ and $\var_{(-)}
\equiv\var(\th-\phi)$, the singularity may be written as
$$
\var(g-q)={1\over4}\var_{(-)}^3-{3\over4}\var_{(+)}^2\var_{(-)}\ ,\qquad
\var(g+q)=-3\var_{(+)}^2+{3\over4}\var_{(+)}\var_{(-)}^2\ \ .
\eqno(newton)$$
The second term in $\var(g-q)$ is actually higher order in the scaling limit,
but we explicitly retain it for comparison purposes below.
We use the scaling {\it ansatz} presented in section 5.1, but the series
in \(orwhat) is modified so that $q\ge1$. We also fix $A=2$, $B=0$,
$C=1$, and $D=1$ in \(map), and set $\hnu/\nu=2$. One finds that
$\nu={1\over 5}$,
$k_1=-h_1=-(s/2)^{{1\over 3}}$, and $k_2=h_2$ where $h_2$ satisfies
$$\eqalign{
0&=t-6h_1{}^2h_2-12h_2{}^2+4\part_t^2h_2\cr
2\part_sh_2&=h_1\part_th_2-h_1\part_sh_1\ \ .\cr}$$
After suitable rescalings and a shift of $h_2$, one finds a scaling function
$\hat{h}=\hat{h}(t+s^{{4\over 3}})$,
which satisfies the Painlev\'e I equation.
This single function determines the universal part of the free energy
as in \(universe).
So the final result is that the free energy is identical to that
found for pure gravity, except with a renormalized topological expansion
parameter, $(t+s^{{4\over 3}})^{-{5\over 2}}$.

We have been unable to generalize this scaling solution to higher
order singularities. While the higher order split singularities all
degenerate at $\phi_c=y^2=1$, for $n>2$ the singularities take the
form
$$
\var(g-q)\propto\var_{(+)}^n\var_{(-)}\ ,\qquad
\var(g+q)\propto\var_{(+)}^n+ b \var_{(+)}^{n-1}\var_{(-)}^2 $$
where $b$ is some constant. These singularities all yield
parametric scaling solutions for square matrices, as described in
section 4. Eq.~\(newton) would have the same form with $n=2$, if it were
not for the appearance of the $\var_{(-)}^3$ term in $\var(g-q)$.
Thus from this point of view, \(newton) seems to be an anomalous case.

\bigskip
We are grateful to Joe Minahan and C. Nappi for valuable communications.
Some of the results in this paper were obtained (i) over two summers
at the Aspen Center for
Physics, (ii) while R.C.M. was visiting the Institute for Theoretical Physics
at U.C.S.B.
and (iii) while V.P. was visiting the Mathematical Sciences Research
Insitute in Berkeley---we thank these institutions for their
hospitality.
This research was supported by NSERC of Canada,
and Fonds FCAR du Qu\'ebec (R.C.M.),
by the NSF under Grant No. PHY89-04035, at UCSB, by the NSF
Grant No. DMS85-05550 at M.S.R.I. (V.P.)
and by DOE Grant No. DE-FG02-90-ER40542 at I.A.S. (V.P.)

\references

\refis{gs} G. Stein, {\it Brim Beauvais}, in {\it The Yale
Gertrude Stein} (ed. R. Kostelanetz), Yale University Press
(New Haven, 1980)

\refis{rdil} R.C. Myers, {\sl Phys. Lett.} {\bf 199B} (1987) 371

\refis{tim2} S. Dalley, C. Johnson, and T. Morris,
`Multicritical complex matrix models and non-perturbative 2-d
quantum gravity', Southampton preprint SHEP-90-91-16 (1991)

\refis{tim} T.R. Morris, {\sl Nucl. Phys.} {\bf B356} (1991) 703

\refis{timstuff} S. Dalley, C. Johnson and T. Morris, `Non-perturbative
two-dimensional quantum gravity', Southampton preprint SHEP-90-91-28 (1991)

\refis{double} C. Bachas and P.M.S. Petropolous, {\sl Phys. Lett.} {\bf
247B} (1990) 363

\refis{simon} S. Dalley, `Critical conditions for matrix models of
string theory', Southampton preprint SHEP-90-91-6 (1990)

\refis{integral} See for example: %
N.I.~Muskhelishvili, {\it Singular Integral Equations
}, P.~Noordhoff N.V. %
(Groningen, Holland, 1953)

\refis{bell} %
G.M.~Cicuta and E.~Montaldi, {\sl Phys. Rev.} {\bf D29} (1984) 1267; %
A.~Barbieri, G.M.~Cicuta and E.~Montaldi, {\sl Nuo. Cim.} {\bf 84A} (1984) %
173; C.M.~Canali, G.M.~Cicuta, L.~Molinari and E.~Montaldi, {\sl Nucl. %
Phys.} {\bf B265} (1986) 485; G.M. Cicuta, L. Molinari, %
E. Montaldi and F. Riva, {\sl J. Math. Phys.} {\bf 28} (1987) 1716

\refis{lafrance} R. Lafrance and R.C. Myers, in preparation

\refis{arley1} A. Anderson, R.C. Myers and V. Periwal, {\sl Phys. Lett.}
{\bf 254B} (1991) 89

\refis{arley2} A. Anderson, R.C. Myers and V. Periwal, {\sl Nucl. Phys.}
{\bf B360} (1991) 463

\refis{joe1} J.A. Minahan, {\sl Phys. Lett.} {\bf 268B} (1991) 29

\refis{joe2joe3} J.A. Minahan, {\sl Phys. Lett.} {\bf 265B} (1991) 362;
University of Virginia preprint UVA-HET-91-07
(1991)

\refis{chihara} T.S. Chihara, {\it Introduction to orthogonal polynomials},
Gordon and Breach (New York, 1978)

\refis{yoneya} S. Nishigaki and T. Yoneya, {\sl Nucl. Phys.} {\bf B348}
(1991) 787

\refis{div} P. Di Vecchia, M. Kato and N. Ohta, {\sl Nucl. Phys.} {\bf
B357} (1991) 495

\refis{thooft} G. 't Hooft, {\sl Nucl. Phys.} {\bf B72} (1974) 461

\refis{bipz} E. Br\'ezin, C. Itzykson, G. Parisi, and J.-B. Zuber,
{\sl Comm. Math. Phys.} {\bf 59} (1978) 35

\refis{ince} E.L. Ince, {\it Ordinary Differential Equations}, Dover
(New York, 1956)

\refis{sade} See for example: %
V.I. Arnold, S.M. Gusein-Zade and %
A.N. Varchenko, {\it Singularities of Differentiable Maps: Volume I}, %
Birkh\"auser (Boston, 1985)

\refis{biz} D. Bessis, C. Itzykson, and J.-B. Zuber, {\sl Adv. Appl.
Math.} {\bf 1} (1980) 109

\refis{kaz} V.A. Kazakov, {\sl Mod. Phys. Lett.} {\bf A4} (1989) 2125

\refis{gross2} D.J. Gross and A.A. Migdal,
{\sl Phys. Rev. Lett.} {\bf 64} (1990) 714

\refis{gross3} D.J. Gross and A.A. Migdal, {\sl Nucl. Phys.} {\bf B340}
(1990) 333

\refis{bre2} E. Br\'ezin, M.R. Douglas, V. Kazakov and S.H. Shenker, Rutgers
preprint RU-89-47 (1989)

\refis{bren} E. Br\'ezin and H. Neuberger, {\sl Phys. Rev. Lett.} {\bf
65} (1990) 2098; {\sl Nucl. Phys.} {\bf B350} (1991) 513

\refis{big3} E. Br\'ezin and V.A. Kazakov, {\sl Phys. Lett.} {\bf B236}
(1990) 144; M.R. Douglas and S.H. Shenker, {\sl Nucl. Phys.} {\bf B335}
(1990) 635; D.J. Gross and A.A. Migdal, {\sl Phys. Rev. Lett.} {\bf 64}
(1990) 127

\refis{doug2} T. Banks, M.R. Douglas, N. Seiberg and S.H. Shenker,
{\sl Phys. Lett.} {\bf B238} (1990) 279

\refis{doug4} M.R. Douglas, N. Seiberg and S.H. Shenker,
{\sl Phys. Lett.} {\bf 244B } (1990) 381

\refis{doug3} M.R. Douglas, {\sl Phys. Lett.} {\bf B238} (1990) 176

\refis{rvec} R.C. Myers, McGill/ITP preprint NSF-ITP-91-69/McGill/91-17,
to appear in {\sl Nuclear Physics B}

\refis{tobe}  To appear

\refis{symp1} R.C. Myers and V. Periwal, {\sl Phys. Rev. Lett.} {\bf 64}
(1990) 3111

\refis{sump} R.C. Myers and V. Periwal, {\sl Phys. Rev. Lett.} {\bf 65}
(1990) 1088

\endreferences
\end

\end